\documentclass[twocolumn]{aastex61}	% Mac OS

\newcommand{\beq}{\begin{equation}}
\newcommand{\eeq}{\end{equation}}
\newcommand{\beqn}{\begin{eqnarray}}
\newcommand{\eeqn}{\end{eqnarray}}

\received{}
\revised{}
\accepted{}
\submitjournal{ApJ}

\shorttitle{BH Formation and Explosion from Rapidly Rotating Very Massive Stars}
\shortauthors{Uchida, H. et al.}

\begin{document}

\title{Black Hole Formation and Explosion from Rapidly Rotating Very Massive Stars}

%\correspondingauthor{Koh Takahashi}
%\email{ktakahashi@astro.uni-bonn.de}
%\author{Koh Takahashi}
%\affil{Argelander-Institut f\"{u}r Astronomie, Universit\"{a}t Bonn,
%D-53121 Bonn, Germany}

\correspondingauthor{Haruki Uchida}
\email{haruki.uchida@yukawa.kyoto-u.ac.jp}
\author{Haruki Uchida}
\affiliation{Center of Gravitational Physics, Yukawa Institute for Theoretical Physics, 
Kyoto University, Kyoto, 606-8502, Japan~} 

\author{Masaru Shibata} 
\affil{Center of Gravitational Physics, Yukawa Institute for Theoretical Physics, 
	Kyoto University, Kyoto, 606-8502, Japan~} 
\affil{Max Planck Institute for Gravitational Physics (Albert Einstein Institute),
	Am M\"uhlenberg 1, Potsdam-Golm 14476, Germany}
\author{Koh Takahashi} 
\affil{Argelander-Institut f\"{u}r Astronomie, Universit\"{a}t Bonn, D-53121 Bonn, Germany}
\author{Takashi Yoshida}
\affil{Department of Astronomy, Graduate School of Science,
	the University of Tokyo, Tokyo, 113-0033, Japan~}

\begin{abstract}
We explore the formation process of a black hole (BH) through the pair-instability collapse of a rotating Population III very massive star 
in axisymmetric numerical relativity. As the initial condition, we employ a progenitor star which is obtained by evolving 
a rapidly rotating zero-age main sequence (ZAMS) star with mass $320M_\odot$ until it reaches a pair instability region. 
We find that for such rapidly rotating model, a fraction of the mass, $\sim 10M_\odot$, forms a torus surrounding 
the remnant BH of mass $\sim 130M_\odot$ and an outflow is driven by a hydrodynamical effect. 
We also perform simulations, artificially reducing the initial angular velocity of the progenitor star, and find
that only a small or no torus is formed and no outflow is driven.  
We discuss the possible evolution scenario of the remnant torus for the rapidly rotating model by considering 
the viscous and recombination effects and show that if the energy 
of $\sim 10^{52}$ erg is injected from the torus to the envelope, the luminosity and timescale of 
the explosion could be of the orders of $10^{43}$ erg/s and yrs, respectively. We also point out the 
possibility for observing gravitational waves associated with the BH formation for the rapidly rotating model 
by ground-based gravitational-wave detectors.
\end{abstract}
\keywords{}

%%%%%%%%%%%%%%%%
% Introduction %
%%%%%%%%%%%%%%%% 

\section{Introduction}
\label{intro}

Gravitational collapse induced by pair instability (PI) is one of the final fates of very massive stars (VMSs) \citep{1967ApJ...148..803R,1967PhRvL..18..379B,1967ApJ...150..131R,1968Ap&SS...2...96F}.
In the carbon-oxygen (CO) core phase, if the CO core enters a region of low density ($\lesssim 10^6~{\rm g/cm^3}$) and high temperature ($\gtrsim 10^9$ K), substantial electron-positron ($e^+ e^-$) pair creation occurs, and then, a volume-averaged adiabatic index decreases below $4/3$. Then, the CO core becomes dynamically unstable and starts gravitational collapse. During the collapse, C and O in the CO core vigorously burn and release enormous {rest-mass energy as thermal energy}.
If the {injected thermal} energy is sufficiently large, it may induce the disruption of the overall progenitor star (PI supernova; PISN) or the ejection for an outer part of the star (pulsational-PISN) \citep{2007Natur.450..390W}.
If the {injected thermal} energy is not large enough to halt the collapse, these VMSs are expected to form black holes (BHs).

Under current theoretical understanding, a star with the mass of a CO core of $\gtrsim130M_\odot$ collapses to a BH after the onset of the pair instability \citep{2002ApJ...567..532H,2002ApJ...565..385U,2016MNRAS.456.1320T,2018ApJ...857..111T}. 
This lower CO core mass limit is estimated to correspond to the zero-age main sequence (ZAMS) mass $M_{\rm ZAMS} \gtrsim 260 M_\odot$  for the case that {there is no efficient (rotational) mixing} with a { sufficiently less efficient mass loss} \citep{2001ApJ...550..372F,2002ApJ...567..532H}. 
The lower ZAMS mass limit may be reduced if the effect of rotation is taken into account in the progenitor evolution because rotational mixing can recycle unprocessed material from the progenitor's outer envelope into the core, and thus, 
finally the more massive CO core may be formed. In the case of rapidly rotating Population III stars, the ZAMS mass limit may reduce to $M_{\rm ZAMS}\gtrsim190M_\odot$ \citep{2012ApJ...748...42C,2012A&A...542A.113Y}.

For metal-rich environments, the formation of a massive star that can form such a high-mass CO core is considered to be unlikely due to efficient wind mass loss \citep{2007A&A...475L..19L,2011MNRAS.412L..78Y,2013MNRAS.433.1114Y,2014MNRAS.438.3119Y}. 
On the other hand, for metal-free environments, it has been speculated that such a massive star can be formed. 
A recent cosmological simulation indicates that $\sim 60\%$ of first stars in number could have $M_{\rm ZAMS}> 240M_\odot$ \citep{2015MNRAS.448..568H}. 
There are two possible reasons for this high percentage. 
The first one is that the typical mass of the first stars can be as large as $\sim 100M_\odot$ due to the {lack of efficient coolants} during its formation \citep[e.g.,][]{2004ARA&A..42...79B}. The second one is that the rate of the line-driven wind mass loss is estimated to be too weak to reduce the total mass during the evolution \citep[e.g.,][]{2009A&A...493..585K}. 

It is pointed out in the simulations of the first-star formation that rapidly rotating massive stars could be formed in a metal-poor environment \citep{2011ApJ...737...75G,2013MNRAS.431.1470S,2015MNRAS.448..568H}.
If a rotating VMS collapses to a BH, there is a possibility that some of the material of the star will form an accretion disk around the BH \citep{2001ApJ...550..372F,2002ApJ...572L..39S}. 
In our previous study for the gravitational collapse of rotating supermassive stars (SMSs) with mass $\gtrsim 10^5M_\odot$ \citep{2017PhRvD..96h3016U}, which is a candidate for seeds of supermassive BHs found in the center of many massive galaxies, we found that if a SMS core is sufficiently rapidly rotating, a surrounding torus with a mass of $\sim 6\%$ of the initial rest mass is formed  after the BH  formation and a fraction of the torus material is ejected as an outflow by a hydrodynamical effect.

If a gravitational collapse of a rotating VMS proceeds in the same way as the SMS, there is a possibility that a torus surrounding a rotating BH is formed and an outflow arises, leading possibly to an observable electromagnetic emission. 
{As a pioneering study, \cite{2001ApJ...550..372F} performed one-dimensional (1D) stellar evolution calculations and axisymmetric (2D) gravitational collapse simulations of rotating VMSs.
For the 2D simulations, they used a Newtonian smoothed particle hydrodynamics code.  
They indicated that for the case of the VMS with the initial mass of $M_{\rm ZAMS} =300M_\odot$ and rotating rigidly with the rotation velocity of $20\%$ of the Kepler rotation at its surface, a He core with mass $\sim 180M_\odot$ is formed and it collapses to a BH. Although they were not able to follow the BH formation in their code, they indicated that a torus would be formed after the BH formation and the mass of the torus and BH would become $\sim30M_\odot$ and $\sim140M_\odot$, respectively.}

For an extension of their study, we perform simulations for the gravitational collapse of rotating Population III VMSs in axisymmetric numerical relativity. {We use a realistic equation of state that includes the contribution from the pressure of gas, radiation, and degenerate electrons including  ${\rm e^+e^-}$ pairs.
For the nuclear reactions, we use the formulation which includes the effect of C, Ne, O and Si burnings and photodissociation reaction including $^4{\rm He} \rightarrow 2{\rm p}+2{\rm n}$.
We also include approximately the effect of neutrino emission.}
The purpose of this paper is to explore the process of gravitational collapse of a rotating VMS  to a BH.  
In particular we pay attention to the properties of the torus and the process of generating an outflow during the formation process of the BH and torus.

This paper is organized as follows. In Section~\ref{setup}, we describe the setup of our numerical simulation. In Section~\ref{result}, we describe the overview of the collapse showing our results of numerical-relativity simulations and describe the processes of the BH formation, torus formation, and generating outflows for models with various rotational velocities. In Section~\ref{discussion}, we discuss the possible evolution scenario of the torus by considering the viscous and recombination effects and estimate the bolometric luminosity of the explosion under the assumption that the energy injection from the torus to the envelope of the progenitor star occurs.  
We also discuss the possibility for observing gravitational waves associated with the BH formation for the rapidly rotating model. 
Section~\ref{conclusion} is denoted to a summary.

\section{Numerical setup}
\label{setup}

\subsection{Stellar evolution phase}
The stellar evolution code described in ~\cite{2016MNRAS.456.1320T,2018ApJ...857..111T} is used to calculate the evolution of zero-metallicity stars. For the initial chemical composition, a result of the Big bang nucleosynthesis \citep{2007ARNPS..57..463S} is used. {47 isotopes are considered in the nuclear reaction network, which is capable of following the main reactions during the hydrostatic evolution of a massive star until the formation of an Fe core.}

Effects of stellar rotation are taken into account by the formulation described in~\cite{2016MNRAS.456.1320T}.
The rotation-induced mixing and angular momentum transport are calculated by using diffusion approximation.
For the calculation of  the diffusion coefficient, we consider {the hydrodynamical instabilities of} the Eddington-Sweet circulation, the Goldreich-Schubert-Fricke instability, the Solberg-H{\o}yrand instability and the dynamical and secular shear instabilities \citep{1989ApJ...338..424P,2000ApJ...528..368H}. 
{At this stage, we consider only the effect of these hydrodynamical instabilities on the transport of angular momentum and do not consider the effects of other additional mechanisms.}

Stellar evolution is calculated from the ZAMS stage until the central temperature, $T_{\rm c}$, reaches ${\rm log}_{10} T_{\rm c}~[{\rm K}]\approx 9.2$ at which the star is unstable to gravitational collapse due to the PI. 
At this stage, we map the resulting 1D stellar evolution models onto 2D grids of axisymmetric gravitational collapse simulations as the initial conditions.

\subsection{Gravitational collapse phase}
For solving Einstein's evolution equations, we use the same method as in ~\cite{PhysRevD.94.021501}. 
We employ the original version of the BSSN~(Baumgarte-Shapiro-Shibata-Nakamura) formalism with a puncture gauge~\citep{1995PhRvD..52.5428S,1999PhRvD..59b4007B,2006PhRvL..96k1101C,PhysRevLett.96.111102}.  
{In the $3+1$ formulation, the metric is defined by the form 
	\begin{equation}
	ds^2 = -\alpha^2 c^2 dt^2  + \gamma_{ij} (dx^i + \beta^i  c dt)(dx^j + \beta^j  c dt),
	\label{metric}
	\end{equation}
	where $c$ is the speed of light and $\alpha,~\beta^i$ and $\gamma_{ij}$ are the lapse function, the shift vector, and the induced metric on three-dimensional (3D) spatial hypersurfaces, respectively. 
	We also define the extrinsic curvature by
	\begin{equation}
	K_{ij}\equiv -\gamma_i^{~\alpha} \gamma_j^{~\beta} \nabla_\alpha n_\beta,
	\label{ext}
	\end{equation}
	where $n^{\mu}$ is a timelike unit-normal vector orthogonal to 3D hypersurfaces. 
	In our BSSN formalism, we evolve $\rho_{\rm g} \equiv ({\rm det}\gamma_{ij})^{-1/6}$,~ $\tilde \gamma_{ij}\equiv \rho_{\rm g}^2 \gamma_{ij}$, 
	$\tilde{A}_{ij} \equiv \rho_{\rm g}^2 (K_{ij}-\gamma_{ij}K^k_{~k}/3)$, $K^k_{~k}$, and $F_i \equiv \delta^{jk}\partial_j \tilde{\gamma}_{ik}$.}
We use the standard 4th-order finite differencing scheme to solve the gravitational-field equations~(see chapter 3 of ~\cite{2016nure.book.....S} for a review). 

{We perform axisymmetric numerical simulations in cylindrical coordinates $(X, Z)$ using a cartoon method for imposing axial symmetry ~\citep{2001IJMPD..10..273A,PhysRevD.67.024033}. A previous work of numerical relativistic 3D simulations of rotating stellar core collapse indicated that nonaxisymmetric deformation during the collapse {did} not occur unless the progenitor star extremely rapidly and differentially rotating ($T_{\rm rot}/|W|\gtrsim 0.01$)~\citep[e.g.,~][]{2005PhRvD..71b4014S}. Here, $T_{\rm rot}$ and $W$ are the initial total rotational kinetic energy and gravitational potential energy, respectively. 
In our models, $T_{\rm rot}/|W|\lesssim 0.003$, and thus, the assumption of axial symmetry is reasonable during the collapse. After the collapse, it is suggested that the formed torus would also deform nonaxisymmetrically due to the Papaloizou-Pringle instability~\citep{1984MNRAS.208..721P}. However, \cite{2011PhRvL.106y1102K} showed that this instability {did} not cause the strong deformation of the torus for the case that its mass is $\sim 10\%$ of the BH mass.}

A nonuniform grid is used for $X$ and $Z$ in the following manner. 
We define the grid spacing at the center by $\Delta X_0 \equiv X_1 -X_0$.　Here $X_0 =0$ and $X_{\rm i}$ is the location of the $i$-th grid. 
We set up the grid spacing for the inside and outside of $X_{\rm in}$ in the following manner: 
for $X_i<X_{\rm in}$, $\Delta X_i \equiv X_i -X_{i-1} = \Delta X_0({\rm const})$, and for $X_{\rm in}\leq X \leq X_{\rm max}$, $\Delta X_i = \eta \Delta X_ {i-1}$,  where $\eta$ is a constant and $X_{\rm max}$ is the size of the computational domain along each axis. $\eta$ determines the nonuniform degree of the grid spacing. 

The grid parameters are set to be $(\Delta X_0,~\eta,~X_{\rm in},~X_{\rm max}) = (0.6R_M,~1.014,~18R_M,~3600R_M)$ until
${\rm log}_{10}T_{\rm c}~[{\rm K}] = 10$, and then, we perform a regrid changing the parameters as $(\Delta X_0,~\eta,~X_{\rm in},~X_{\rm max}) = (0.005R_M,~1.014,~0.4R_M,~2000R_M)$. Here, $R_M\equiv GM/c^2 $ where $G$ and $M$ are the gravitational constant and the total mass of the progenitor star at the end of the stellar evolution calculation, respectively. For the model of this paper, $M\approx290M_\odot$ and $R_M\approx 4.3\times 10^7~{\rm cm}$ (see Section~\ref{models}). To confirm that numerical results with different grid resolutions agree reasonably  with each other, we also perform a simulation with the low-resolution grid parameters such that $(\Delta X_0,~\eta,~X_{\rm in},~X_{\rm max})=(0.007R_M,~1.014,~0.4R_M,~1700R_M)$ for the most rapidly rotating model (model a10, see Section \ref{models}). We show that the numerical results agree reasonably with each other for two resolution models in Section~\ref{BH}.

{We use one of \cite{2000ApJS..126..501T} equations of state which includes the contribution from radiation, ions as ideal gas, electrons, positrons and corrections for Coulomb effects. For electrons and positrons, the relativistic effect, the effect of degeneration and electron-positron pair creation are taken into account.}

In order to include the energy generation by thermonuclear reactions and photodissociation reactions including $^4{\rm He} \rightarrow 2{\rm p}+2{\rm n}$, the nuclear reaction calculation is divided into the following two categories.

\begin{figure}[htbp]
	\includegraphics[width=1\linewidth]{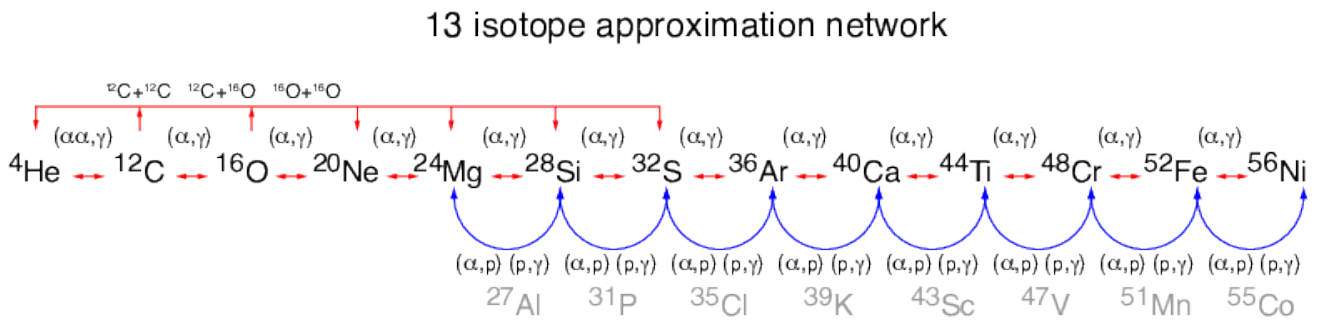}
	\caption{Isotopes included in the "approx 13" nuclear reaction network code. This figure is taken from ${\rm http://cococubed.asu.edu/code\_ pages/burn\_ helium.shtml}$.}
	\label{fig:isotope13}
\end{figure}
For $T<5\times 10^9$ K, we use the "approx13" nuclear reaction network code that includes the $13~\alpha$-chain elements \citep{1999ApJS..124..241T}. Figure~\ref{fig:isotope13} shows the $13~\alpha$-chain elements included in this code.
On the other hand, for $T\geq 5\times 10^9$ K, we assume that the abundances of isotopes are in a state of nuclear statistical equilibrium (NSE), and solve NSE equations including protons, neutrons and $13~\alpha$-chain elements. In our calculation, we neglect the electron capture process, and thus, 
we assume the electron fraction $Y_{\rm e}=0.5$. 
We include the thermal effect of the nuclear reactions by using the same formulation as we described in  ~\cite{2017PhRvD..96h3016U}.

In our calculation, the thermal neutrino energy loss is approximately included in the right-hand side of the equations of motion as
\begin{equation}
\nabla_\lambda T^{\lambda \mu} =\frac{u^\mu \rho}{c} {q}_{\rm neu}.
\label{neu}
\end{equation}
Here, $T_{\mu \nu},~u^\mu,~\rho$, and ${q}_{\rm neu}$ are the energy momentum tensor, four velocity, rest-mass density of fluid and thermal neutrino emission rate, respectively. To calculate ${q}_{\rm neu}$, we use the analytic fitting formulas introduced in ~\cite{1996ApJS..102..411I}. 
Because neutrinos are optically thin for most regions of the collapsing star and remnant torus, we assume that neutrinos are entirely optically thin in this approximation. 
{In reality, just before the BH formation, the matter density of the central region of the collapsing star exceeds  $10^{11}~{\rm g/cm^3}$ and neutrinos would become optically thick. However, our simulations show that this region immediately falls into the BH. Hence, it would be safe to consider that the formulation of Equation~(\ref{neu}) is applicable for our present study (see Section \ref{before} for more detail).}

We confirmed that our 2D code successfully reproduces 1D results for a PISN calculation of a relatively low-mass VMS (see Appendix~\ref{A}). This illustrates that with our prescription, the effect of key nuclear burning in the self-gravitating system is correctly taken into account.

\subsection{Models}
\label{models}
We select a progenitor ZAMS star with its initial mass $M_{\rm ZAMS} =320M_\odot$. 
For its initial rotation profile, we employed the rigid rotation with the rotation velocity of $50\%$ of the Kepler rotation at its surface (the angular velocity and radius of the star are $\approx 2.1\times 10^{-4}~{\rm s}^{-1}$ and $\approx 5.9\times 10^{13}~{\rm cm}$, respectively). As we already mentioned, we perform a stellar evolution calculation until ${\rm log}_{10} T_{\rm c}~[{\rm K}]$ reaches $\approx 9.2$. 

\begin{figure}[t]
	\includegraphics[scale=0.7]{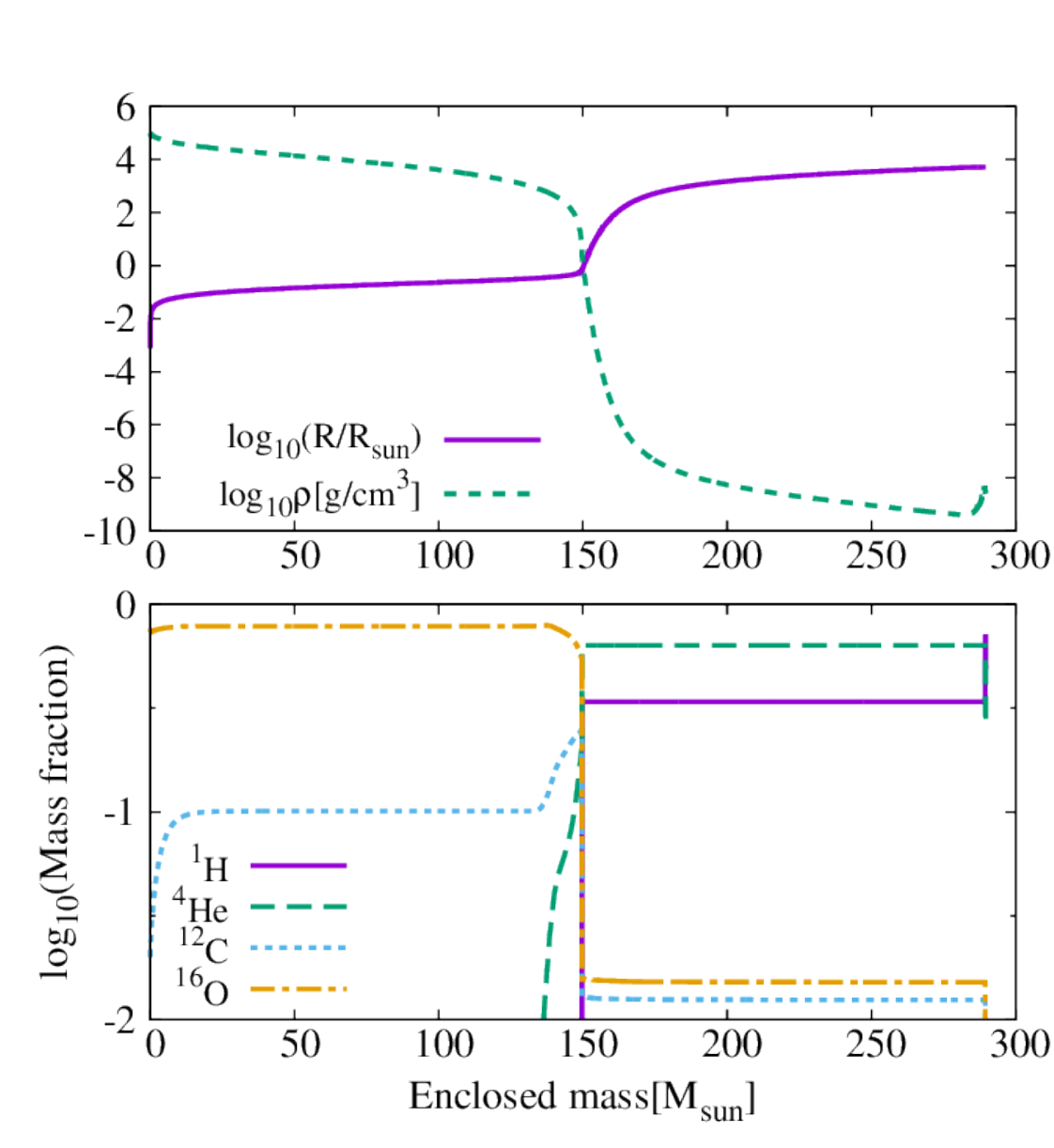}
	\caption{Progenitor profile of our model as functions of the enclosed mass at the end of the stellar evolution calculation. Upper panel shows the radius (solid curve) and density (dashed curve) profiles, respectively. Lower panel shows the chemical distribution. The solid, dashed, dotted and dashed-dotted curves show the chemical abundance of $^1$H, $^4$He, $^{12}$C and $^{16}$O, respectively.  }
	\label{fig:prog}
\end{figure}

Figure~\ref{fig:prog} shows the profiles of the density and radius (upper panel) and chemical distribution (bottom panel) of the progenitor star at the end of the stellar evolution calculation. The region of $\lesssim150M_\odot$ is a compact core composed mainly of C and O and $\gtrsim 150M _\odot$ is a broaden envelope composed mainly of H and He. During the evolution, the star loses a  part of its envelope due to the mass loss, which is enhanced by the rotation \citep{1998A&A...329..551L,2000A&A...361..159M}. The total mass of the star at this end stage is $M\approx 290M_\odot$ and the outer edge of the CO core is located at a radius of {$\approx 4.3\times 10^{10}~{\rm cm}~(\approx 1000R_M)$}.
\begin{figure}[t]
	\includegraphics[scale=0.7]{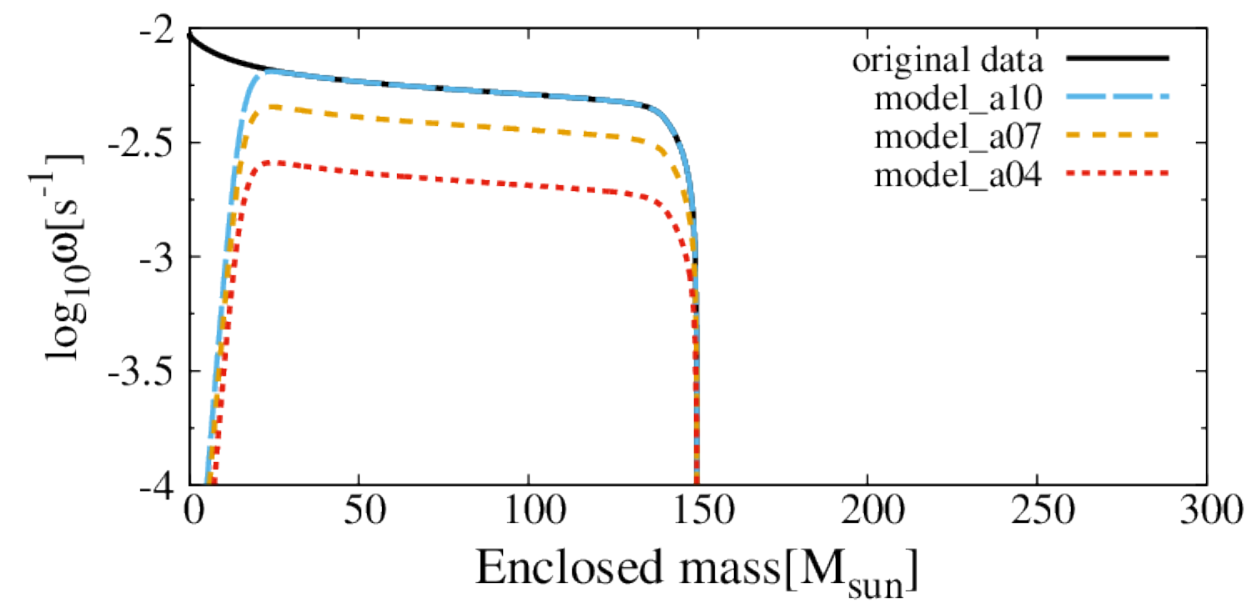}
	\caption{The angular velocity profile for the original data (solid curve), models a10 (long-dashed curve), a07 (dashed curve) and a04 (dotted curve).}
	\label{fig:progomg}
\end{figure}

{The angular velocity profile, $\omega$, of this progenitor star is denoted by the black solid curve in Figure~\ref{fig:progomg}. } {The CO core approximately has a uniform rotation profile due to the convection that occurs in the entire He core during the He core burning phase. }
On the other hand, the envelope rotates approximately rigidly at a very small angular velocity of the order of  $10^{-10}~{\rm s}^{-1}$. This is due to the fact that during the late stage of the stellar evolution, the convection occurs entirely in the envelope for this model, redistributing the angular momentum, and the rigid rotation state is achieved. 
Because of the {expanded} structure of the envelope (with the radius of its surface $\approx 3.6 \times 10^{14}$ cm), the angular velocity is much smaller than that of the core.

{If additional angular momentum transport mechanisms work efficiently during the stellar evolution phase, the final core angular velocity would be reduced. For example, \cite{2018ApJ...857..111T} performed 1D stellar evolution calculations of rotating VMSs with and without the effect of the magnetic stress modeled by the  Tayler-Spruit dynamo (TS dynamo, \cite{2002A&A...381..923S}). {They indicated that the magnetic models have 10 times slower rotation rate than the non-magnetic models.} To consider the case for which the angular velocity is decreased due to such effects,} we simulated two additional models for which the angular velocity is multiplied by a factor of $0.7$ (model a07) and $0.4$ (model a04) together with the original model (model a10).

At the start of simulations, we artificially reduce the angular velocity within a small central region of enclosed mass $\lesssim 30M_\odot$ for all the models. 
This is because the angular velocity profile for the original data sharply 
rises at the center, and hence, the dimensionless spin parameter of the BH  in the early stage of the collapse ($M_{\rm BH} \lesssim 30M_\odot$) becomes too large ($q_{\rm BH} \approx 1$), and numerical accuracy deteriorates at this time. Here, $M_{\rm BH}$ and $q_{\rm BH}$ are the mass and dimensionless spin parameter of the BH, respectively. To avoid this difficulty, we initially reduce the angular velocity within the region of mass $\lesssim 30M_\odot$ for all the  models.  Since the total angular momentum of this inner region is only $\approx 1\%$ of the whole core, this handling would not affect strongly the properties of the final BH, torus and outflow described in Section~\ref{result}.
The long-dashed, dashed and dotted curves of Figure~\ref{fig:progomg} show the angular velocity distribution for models a10, a07 and a04, respectively. 

{Before closing this section, we predict the mass and spin of the BH formed after the gravitational collapse of the core in these models. First, we calculate the quantities of the BH for the hypothetical case that the whole core collapses to form a BH. Then the dimensionless spin parameter of the BH is estimated by {$q_{\rm BH}\approx cJ_{\rm core}/GM^2_{\rm core}$} where $J_{\rm core}$ and $M_{\rm core}$ are the total angular momentum and mass of the core, respectively (for the original data, $J_{\rm core}\approx 2.2\times 10^{53}~{\rm g~cm^2~s^{-1}}$).　Substituting $M_{\rm core} =150M_\odot$, we find that $q_{\rm BH}$ for models a10, a07 and a04 are $1.1,~0.78$ and $0.44$, respectively. 
For model a10, $q_{\rm BH}$ exceeds 1. This suggests that all the fluid elements of the core would not collapse into a BH and that some elements are likely to form a torus around the BH for model a10. 

Next, we analytically estimate $M_{\rm BH}$ and $q_{\rm BH}$ for all the models {based on} the method described in ~\cite{0004-637X-818-2-157}. 
{We describe briefly the method of the estimation. We assume that a seed BH is formed at the center of the collapsing VMS core and it dynamically grows while sequentially absorbing fluid elements from lower values of specific angular momentum, $j$. Then, we calculate approximately the mass, $m(j)$, and spin, $a(j)=cJ(j)/Gm(j)$ of the hypothetically growing BH at each moment by
\begin{equation}
m(j) = \int dV \rho \Theta( j-X^2 \omega)  ,
\end{equation}
and
\begin{equation}
J(j) = \int dV \rho X^2 \omega \Theta(j-X^2 \omega) ,
\end{equation}
where $\Theta(x)$ is the step function which satisfies $\Theta(x)=1~(x\ge0)$ and $\Theta(x)=0~(x<0)$. 
Here, we neglect all the relativistic corrections in this analysis because they only give a minor contribution. 
Assuming that the BH is a Kerr BH, we can calculate $j_{\rm ISCO}(j)$ by inserting $m(j)$ and $a(j)$ to Equation~(2.21) of ~\cite{1972ApJ...178..347B}. Here $j_{\rm ISCO}$ is the specific angular momentum, which is needed for a test particle to rotate at an innermost stable circular orbit (ISCO) in the equatorial plane around the BH. 
We assume that the growth of the BH would terminate at the moment at which  $j$ becomes larger than $j_{\rm ISCO}(j)$.}
Then, we find that $(M_{\rm BH},~q_{\rm BH})$ for models a10, a07 and a04 are  $(122M_\odot,~0.83),~(147M_\odot,~0.72)$ and $(150M_\odot,~0.44)$, respectively. 
We expect that the remaining mass of the core will form a torus around the BH or be ejected as an outflow at the formation of a torus. For model a04, we find that all the fluid elements of the core are likely to form a BH.}

\section{Result}
\label{result}
\subsection{Before the BH formation}
\label{before}

{First, we briefly describe the effect of nuclear reactions and neutrino emission until the BH  formation. 
Although the time taken from the start of the gravitational collapse until the BH formation is different for each model due to the difference of the centrifugal force strength, the qualitative feature of the collapse dynamics depends weakly on the angular momentum for each model. Thus, we focus only on model a10.
Hereafter, we take the time of the BH formation as the origin of time. }

{Figure~\ref{fig:nuc_a10} shows the time evolution of the central temperature (solid curve), total neutrino emission rate, $Q_{\nu}$ (dashed curve), and total energy generation rate of the nuclear reactions, $Q_{\rm nuc}$ (dotted curve), until the BH formation for model a10. 
$Q_{\nu}$ and $Q_{\rm nuc}$ are defined by}
 \begin{equation}
Q_{\nu} = \int \rho_* {q}_{\rm neu}\alpha  d^3 x,
\end{equation}
and
\begin{equation}
Q_{\rm nuc} = \int \rho_* {q}_{\rm nuc}\alpha  d^3 x,
\end{equation}
{respectively. Here, ${q}_{\rm nuc}$ is the energy generation rate of the nuclear reactions and $\rho_* \equiv \rho u^t \sqrt{-g}$. Until $T_{\rm c}\lesssim 5\times 10^9$ K ($t\lesssim-6$ s), the gravitational collapse is  {decelerated by the energy generation due to the nuclear burning}.
However, since our star model is massive enough, the collapse cannot be halted by the energy injection from the nuclear burning. 
Then, when $T_{\rm c}$ reaches $\sim 5\times 10^9$ K ($t\gtrsim-6$ s), partial photodissociation reaction of heavy elements into $^4$He occurs at the center, and thus, the total energy generation rate of the nuclear reactions becomes a negative value. 
When $T_{\rm c}$ reaches $\sim 10^{10}$ K ($t\gtrsim-0.5$ s), photodissociation reaction of $^4$He sets in at the center. 
Since this reaction increases the number density of nucleons, the gas pressure increases sharply at the center, and then, the adiabatic index is slightly increased. However, this effect does not play a significant role for halting the collapse. 
 \begin{figure}[t]
	\includegraphics[scale=0.7]{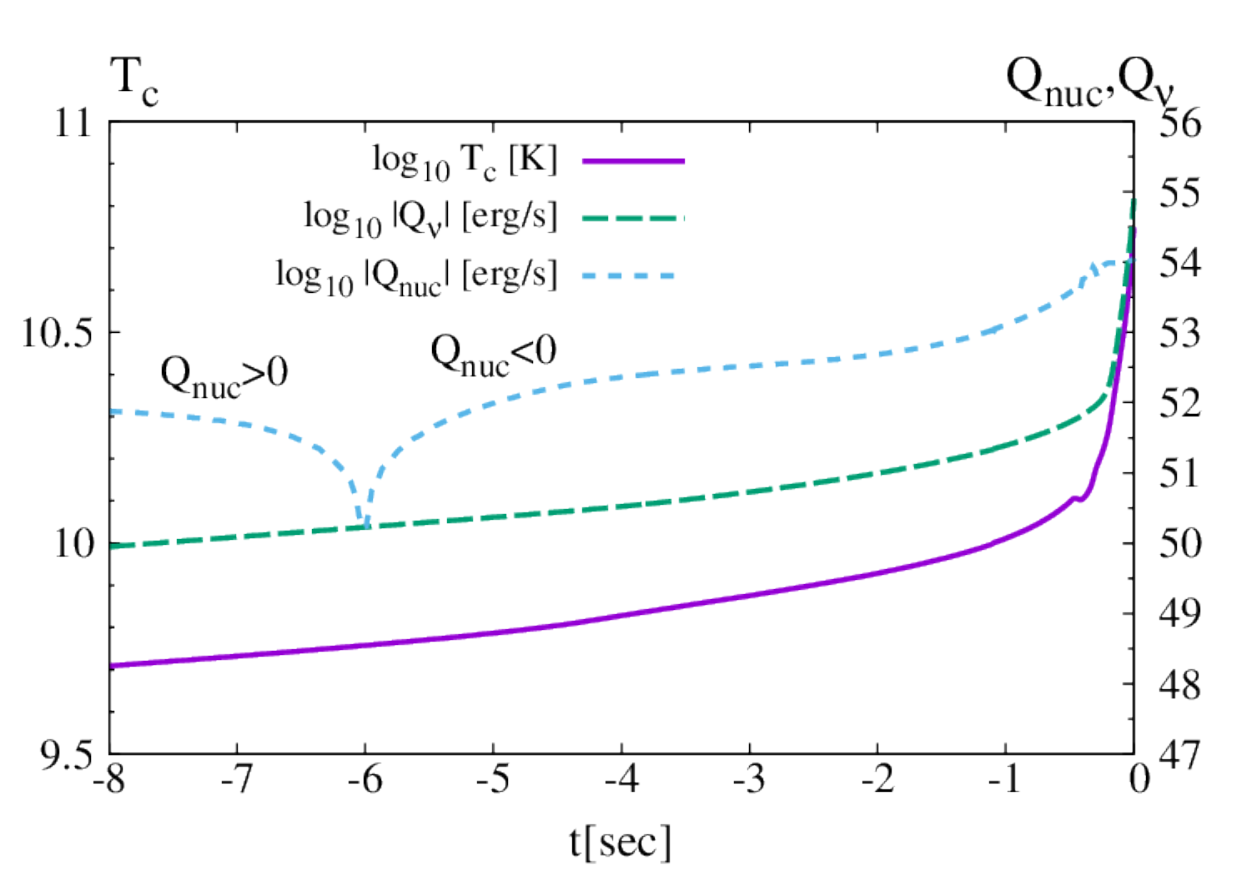}
	\caption{Time evolution of several quantities until the BH formation for model a10. The solid, dashed and dotted curves show the central temperature, total neutrino emission rate and total energy generation rate of the nuclear reactions, respectively. 
		For $t\lesssim -6$ s, $Q_{\rm nuc}>0$ and for $t\gtrsim -6$ s, $Q_{\rm nuc}<0$. Here, $t=0$ is the time at which a BH is formed.}
	\label{fig:nuc_a10}
\end{figure}
	
In our formulation, neutrinos take away a large amount of the thermal energy of the central region, and thus, the central region collapses promptly to a BH. In reality, neutrinos in the high density region with $\rho \gtrsim 10^{11}~{\rm g/cm^3}$ would become optically thick, and thus, the neutrino cooling would not be efficient in this region. 
In order to investigate this effect, we also performed a simulation for the same condition as model a10 but not including the neutrino emission (this corresponds to assuming that all neutrinos are trapped). We find that although the high density core supported by the gas pressure is temporary formed and the collapse is delayed by $\sim 0.3$ s, the core immediately collapses to a BH and our final results do not change qualitatively.

\subsection{After the BH formation}
\label{after}

\begin{figure*}[htbp]
	\begin{minipage}{0.3\hsize}
		\begin{center}
			\hspace*{-5em}  
			\includegraphics[width=1\linewidth,bb=0 0 576 403,angle=0,trim=120 20 120 60]{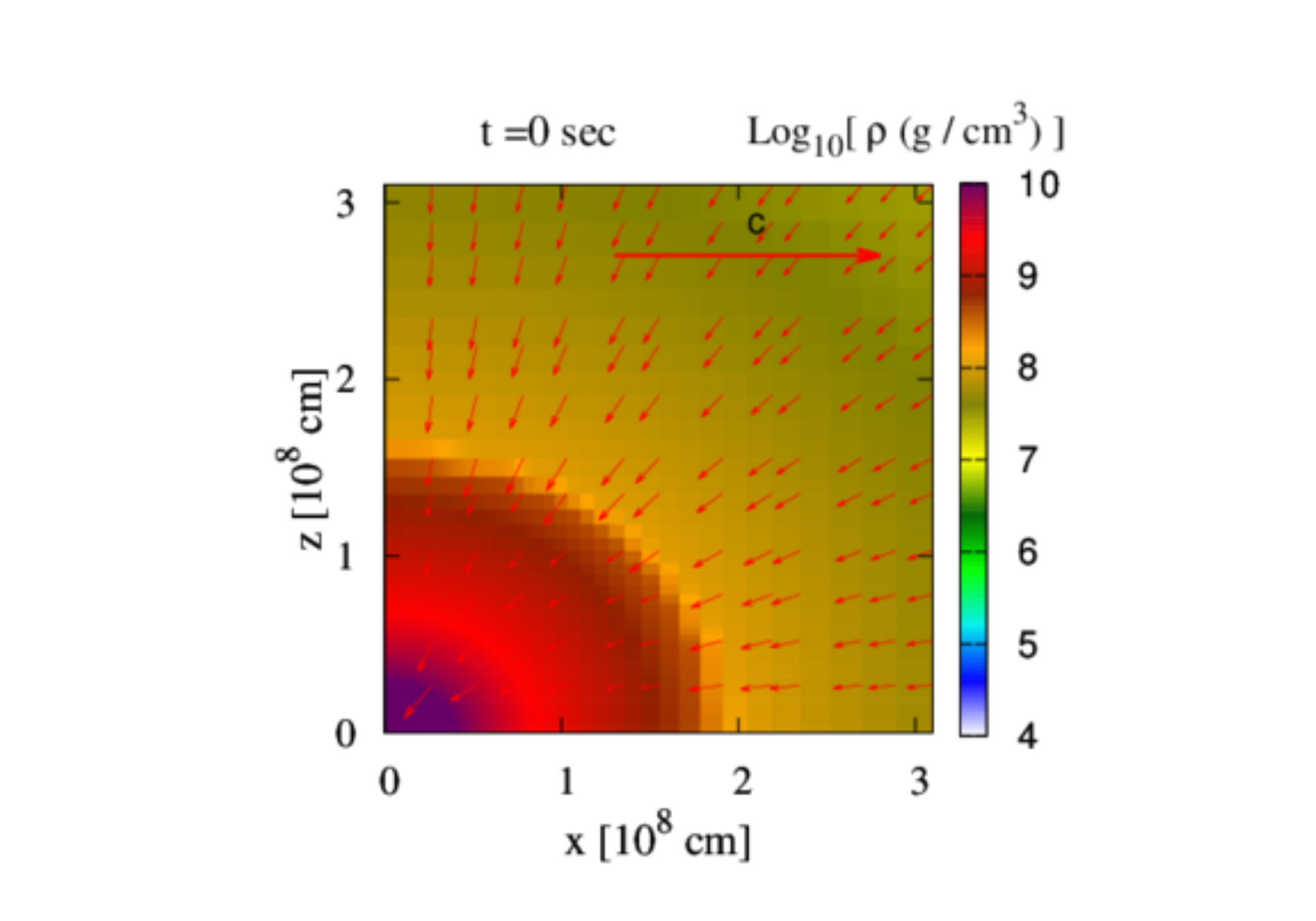}
		\end{center}
	\end{minipage}
	\begin{minipage}{0.3\hsize}
		\begin{center}
			\hspace*{-5em} 
			\includegraphics[width=1\linewidth,bb=0 0 576 403,angle=0,trim=120 20 120 60]{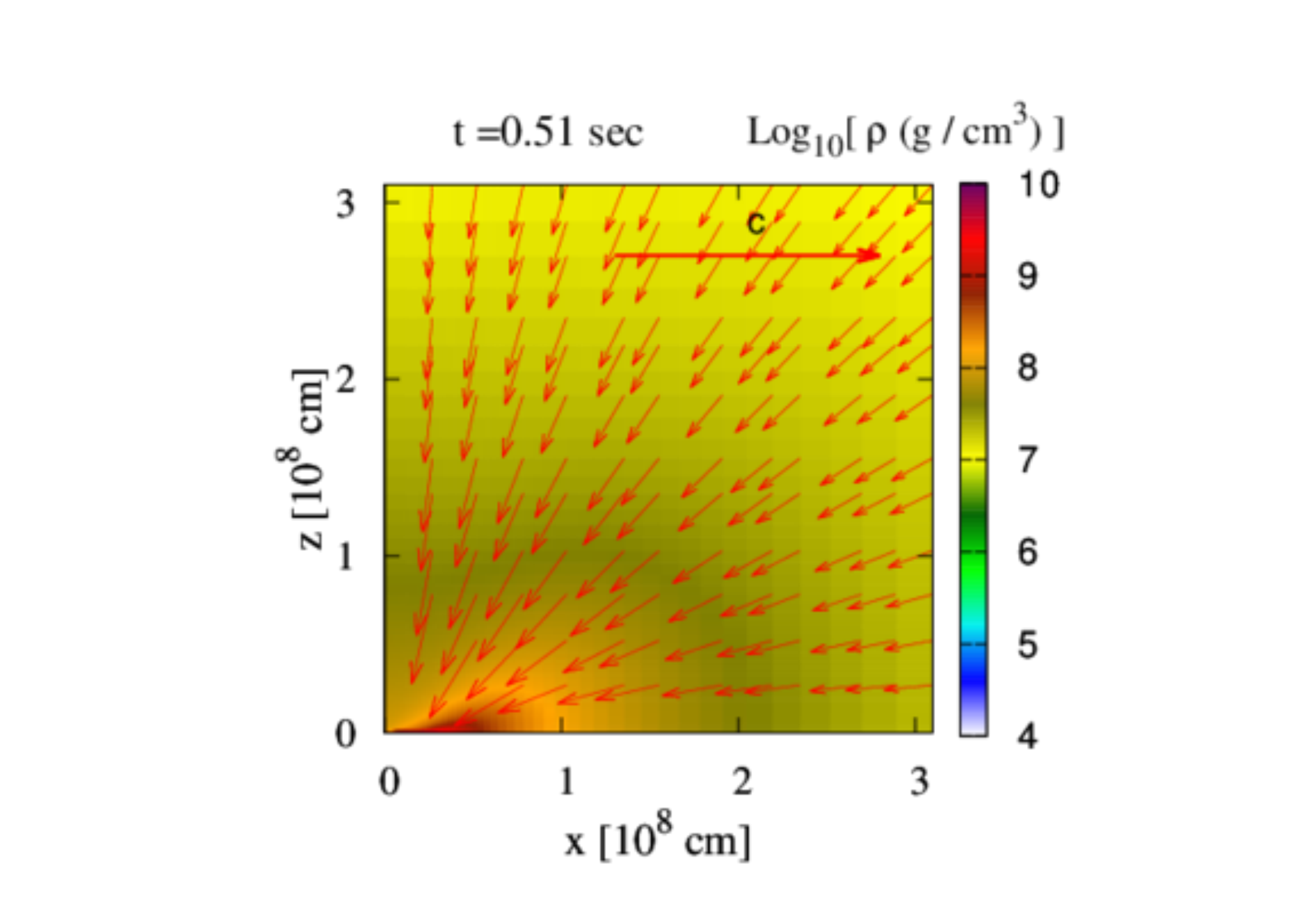}
		\end{center}
	\end{minipage}
	\begin{minipage}{0.3\hsize}
		\begin{center}
			\hspace*{-5em} 
			\includegraphics[width=1\linewidth,bb=0 0 576 403,angle=0,trim=120 20 120 60]{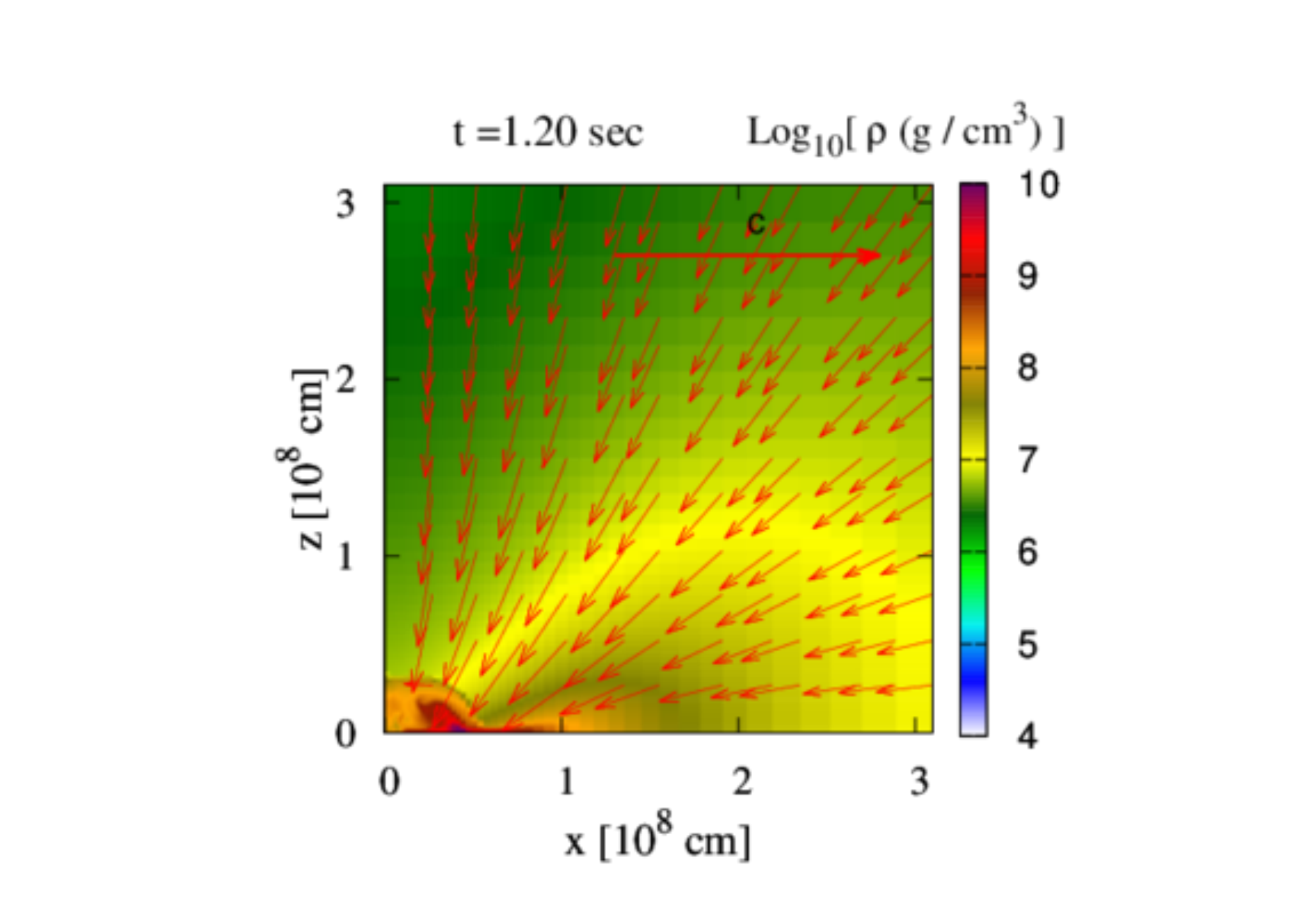}
		\end{center}
	\end{minipage}
%--------------------------------------------------------------
	\begin{minipage}{0.3\hsize}
	\begin{center}
			\hspace*{-5em} 
			\includegraphics[width=1\linewidth,bb=0 0 576 403,angle=0,trim=120 20 120 30]{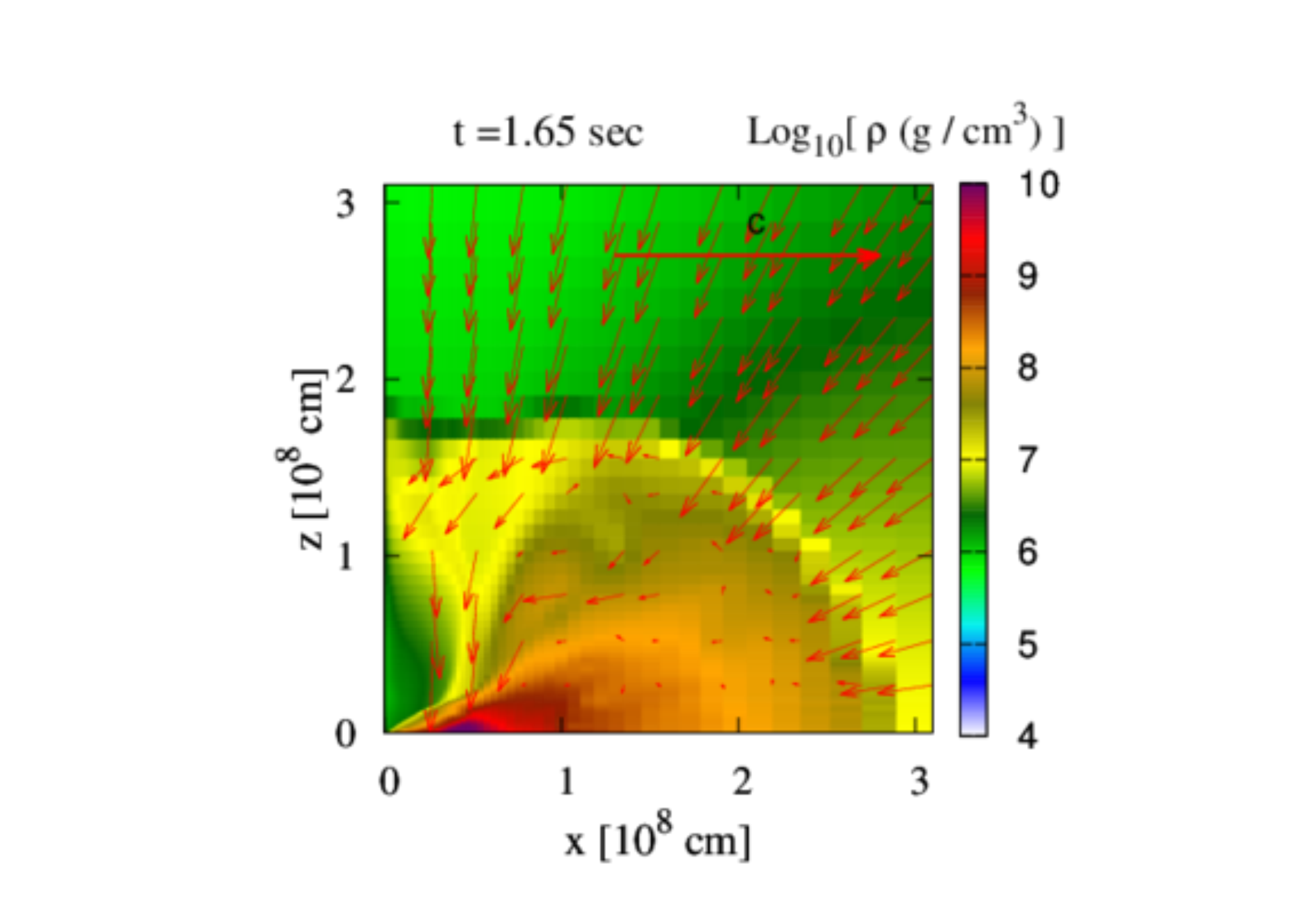}
	\end{center}
	\end{minipage}
	\begin{minipage}{0.3\hsize}
		\begin{center}
			\hspace*{-5em} 
			\includegraphics[width=1\linewidth,bb=0 0 576 403,angle=0,trim=120 20 120 30]{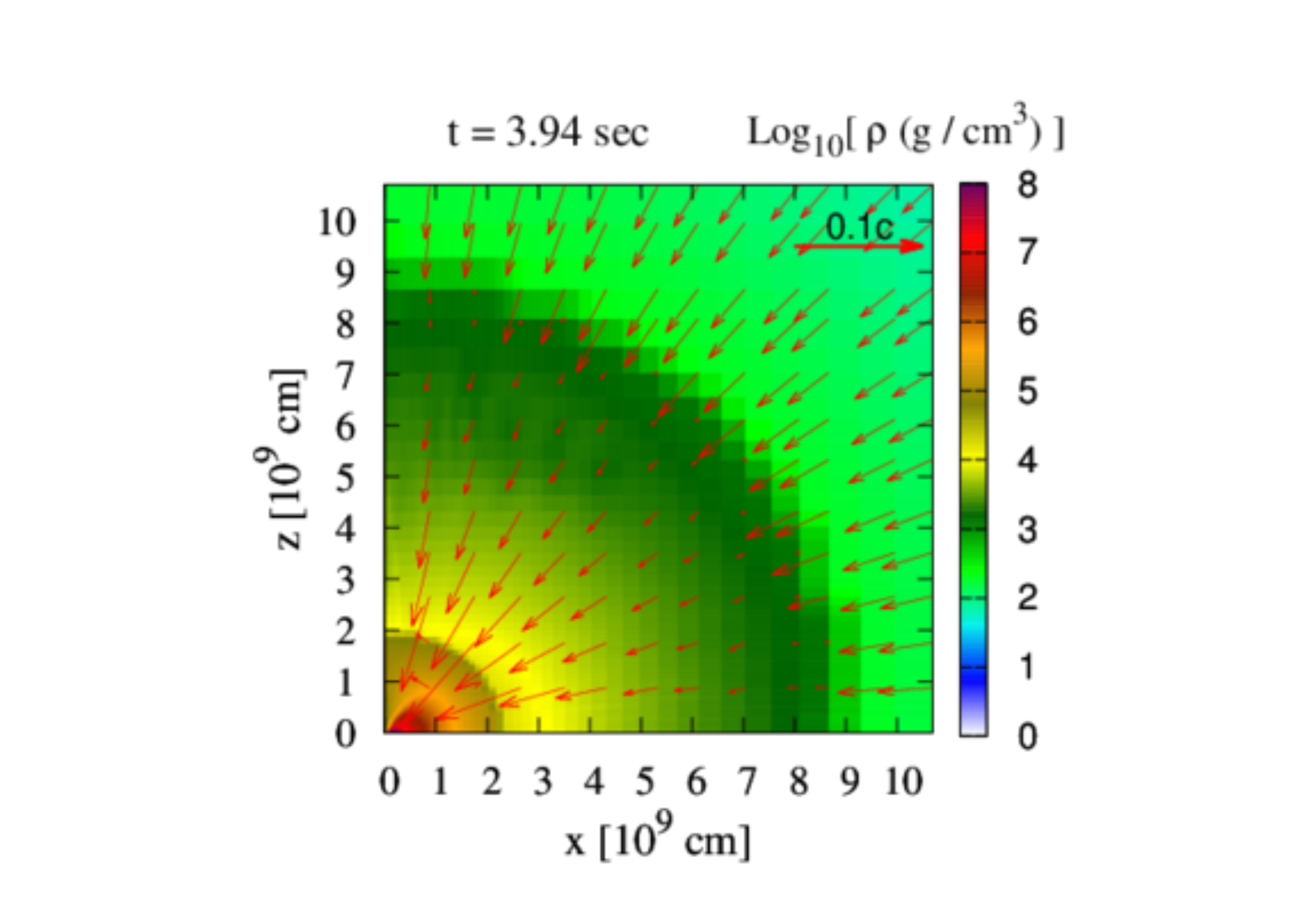}
		\end{center}
	\end{minipage}
	\begin{minipage}{0.3\hsize}
		\begin{center}
			\hspace*{-5em} 
			\includegraphics[width=1\linewidth,bb=0 0 576 403,angle=0,trim=120 20 120 30]{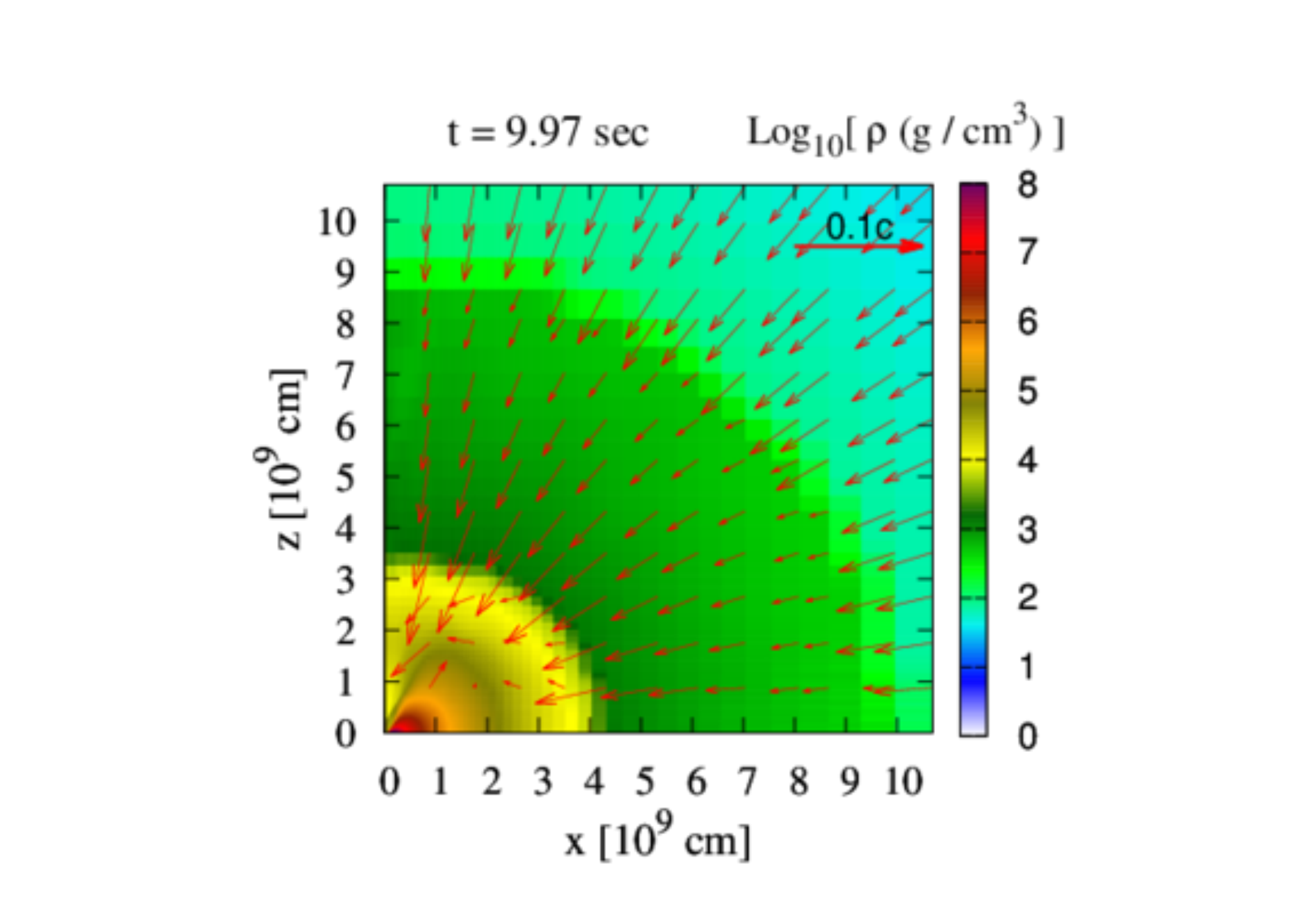}
		\end{center}
	\end{minipage}
%--------------------------------------------------------------
	\begin{minipage}{0.3\hsize}
		\begin{center}
			\hspace*{-5em} 
			\includegraphics[width=1\linewidth,bb=0 0 576 403,angle=0,trim=120 20 120 30]{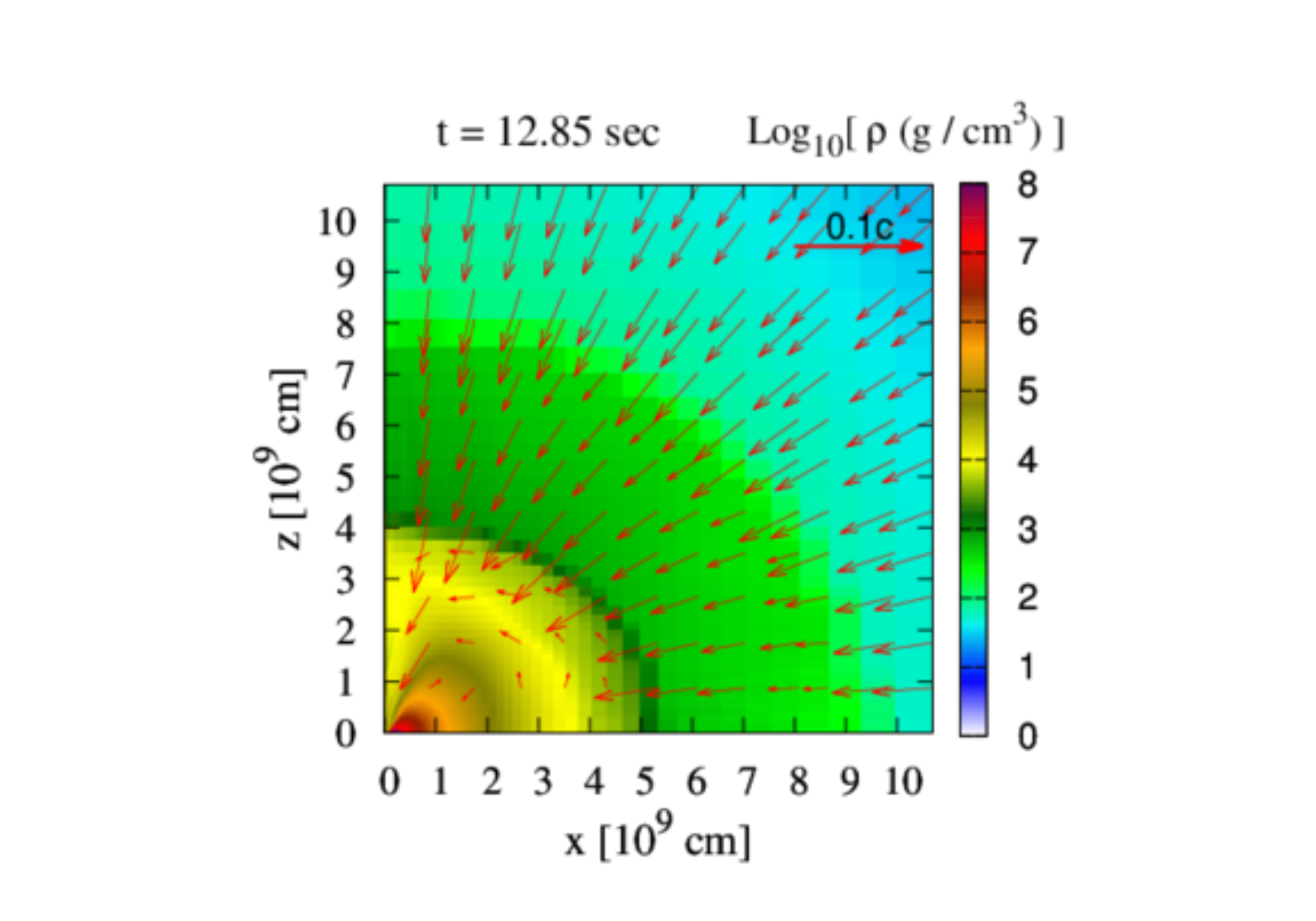}
		\end{center}
	\end{minipage}
	\begin{minipage}{0.3\hsize}
		\begin{center}
			\hspace*{-1em} 
			\includegraphics[width=1\linewidth,bb=0 0 576 403,angle=0,trim=120 20 120 30]{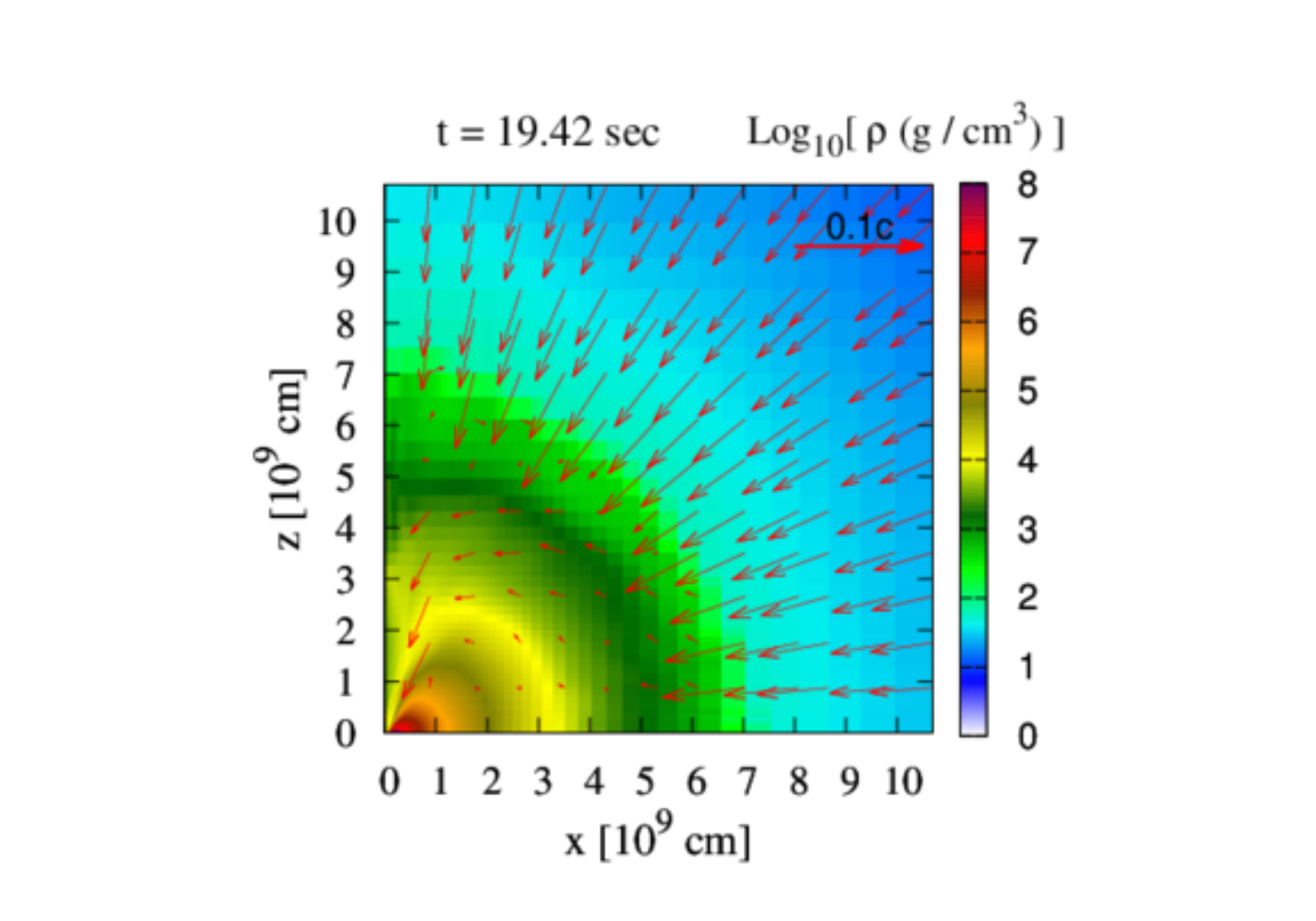}
		\end{center}
	\end{minipage}
	\begin{minipage}{0.3\hsize}
		\begin{center}
			\hspace*{1.5em} 
			\includegraphics[width=1\linewidth,bb=0 0 576 403,angle=0,trim=120 20 120 30]{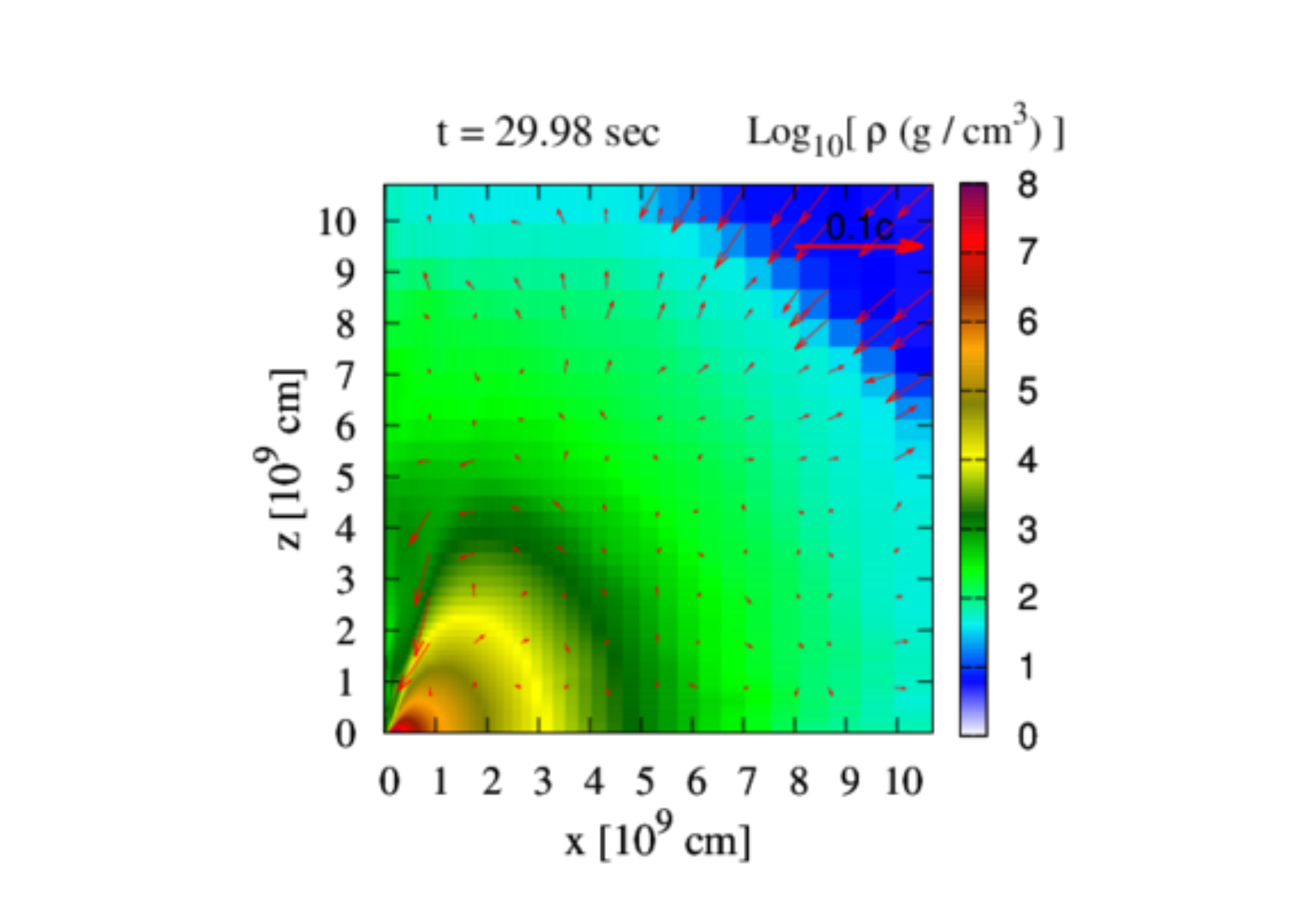}
		\end{center}
	\end{minipage}
	\caption{Snapshots of the density profiles during the VMS core collapse for model a10. 
		The origin of time is taken at the time of the BH formation.
		The red arrows denote the velocity profile, $u^i/u^t(i=X,Z)$, which are normalized by $c$ or $0.1c$ for which the size is indicated in the upper right-hand corner of each snapshot. 
		The 5th--9th panels show zoom-out views of the outer region}
	\label{fig:collapse}
\end{figure*}

 \begin{figure}[htbp]
	\includegraphics[scale=0.7]{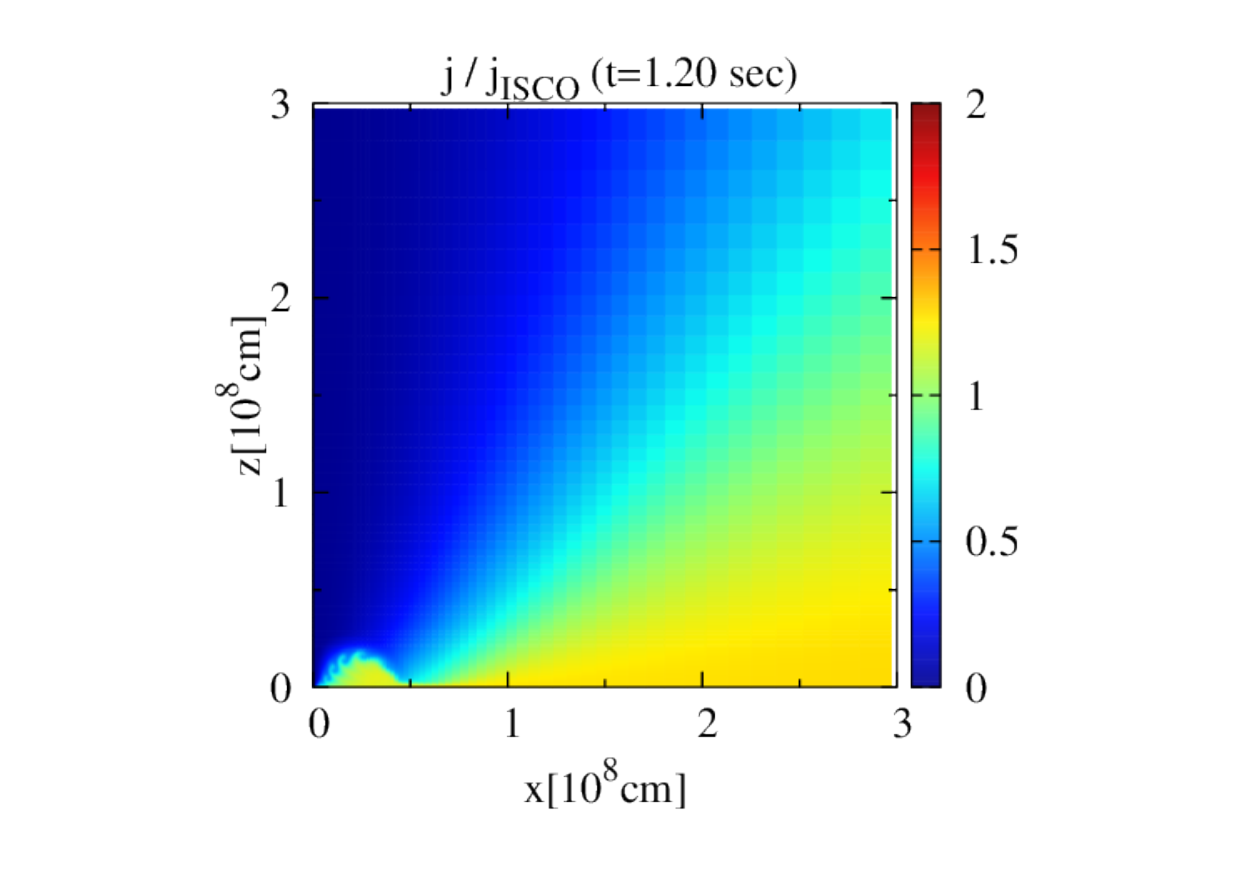}
	\caption{The distribution of $j/j_{\rm ISCO}$ of the torus for model a10 at $t=1.20$ s. Here, $j$ is the specific angular momentum of the fluid elements and $j_{\rm ISCO}$ is the specific angular momentum, which is needed for a test particle to rotate at an ISCO in the equatorial plane around the BH. }
	\label{fig:j2d}
\end{figure}

{Although the qualitative feature of the collapse dynamics until the BH formation depends weakly on it,
the final outcomes depend strongly on the initial rotation velocity.} In this section, we outline the process of the gravitational collapse after the BH formation for each model separately.

\subsubsection{model a10}
Figure \ref{fig:collapse} displays the snapshots of the density profile after the BH formation for model a10. 
As already mentioned,  the collapse cannot be halted by the energy injection from the nuclear burning, and thus, a BH is formed at the center without any strong bounce (1st panel of Figure~\ref{fig:collapse}). 
Indeed, the central density exceeds the nuclear density ($\sim10^{14}~{\rm g/cm^3}$) just before the BH formation ($t\gtrsim -5$ ms), and thus, there is a possibility that a proto-neutron star (PNS) is formed temporarily at the center. However, this is not the case because  the VMS which we consider here is massive enough.  \cite{2007ApJ...666.1140N} shows that for the case of the gravitational collapse of mostly isentropic iron core with $s_{\rm i}>7.5k_{\rm B}$ (core mass $\gtrsim 10M_\odot$), the core collapses promptly to a BH without a quasi-stationary PNS formation. Here, $s_{\rm i}$ and $k_{\rm B}$ are the initial specific entropy of the core and Boltzmann constant, respectively. 
Because the cores of our models are mostly isentropic and $s_{\rm i}\approx 15k_{\rm B}$, it is safe to ignore the effect of the PNS formation for our models. 

Since the fluid elements conserve their specific angular momentum, $j$, the growth of the BH is suppressed when their specific angular momentum is larger than $j_{\rm ISCO}$ for the growing BH. Here, 
\begin{equation}
j\equiv c h u_{\varphi},
\end{equation}
where $h$ and $u_{\varphi}$ are the specific enthalpy and the azimuthal component of the four velocity for the fluid, respectively. Fluid elements with $j>j_{\rm ISCO}$ form a torus surrounding the BH (2nd panel of Figure~\ref{fig:collapse}). 

At the same time, a fraction of the fluid elements in the torus is pushed inward by the inertia of the entire torus matter which has small infall velocity. Then, a part of material falling from a high latitude hits the inner part of the torus, and then, due to the strong encounter among the fluid elements, shocks are formed. As a result, a dense bubble (of a torus shape) is formed by the shock heating near the inner edge of the torus ($X\lesssim 5\times 10^7~{\rm cm}$, 3rd panel of Figure~\ref{fig:collapse}). 

{Figure~\ref{fig:j2d} displays the distribution of $j/j_{\rm ISCO}$ at $t=1.20$ s.  $j_{\rm ISCO}$ is calculated in the following manner. 
	At $t=1.20$ s, the mass and spin of the BH are $M_{\rm BH}\approx 112M_\odot$ and $a_{\rm BH}\approx 0.84 M_{\rm BH}$, respectively. Assuming that the BH is a Kerr BH, we calculate $j_{\rm ISCO}$ by inserting $M_{\rm BH}$ and $a_{\rm BH}$ to Equation~(2.21) of ~\cite{1972ApJ...178..347B}.   
	Since a majority of the matter in this bubble has larger values of $j$ than $j_{\rm ISCO}$ and very hot, it does not accrete onto the BH and starts expanding as an outflow forming shocks (4th and 5th panels of Figure~\ref{fig:collapse}). }

 The expansion of the shock front decelerates temporarily at ($3$--$5)~\times 10^9$ cm away from the BH because the pressure behind the shock and the ram pressure of the infalling matter are balanced at this point (6th panel of Figure~\ref{fig:collapse}). However, in a few seconds afrer the stalling of the shock, the expansion of the shock front sets in again because of the decrease of the density of the infalling matter and resulting decrease of the ram pressure (7--9th panels of Figure~\ref{fig:collapse}). {At $t=30$ s, the shock front is located at a radius of $\approx 10^{10}~{\rm cm}$ and its expansion velocity is $\approx 5\times 10^8$ cm/s. We estimate that the total energy of outflows injectable into the envelope is of the order of $10^{50}$ erg.}
 
We define the total mass, $M_{>6}$, total internal energy, $E_{>6}$, and total neutrino emission rate, $Q_{\nu,>6}$, of the region with $\rho \geq 10^6~{\rm g/cm^3}$. These quantities represent approximately the quantities of the torus. $M_{>6},~E_{>6}$ and $Q_{\nu,>6}$ are calculated by
 \begin{equation}
 M_{>6}= \int_{\rm \rho \ge 10^6 g cm^{-3}} \rho_* d^3 x,
 \end{equation}
 \begin{equation}
 E_{>6}= \int_{\rm \rho \ge 10^6 g cm^{-3}} \rho_* \epsilon \, d^3 x,
 \end{equation}
 and
 \begin{equation}
 Q_{\nu,>6} = \int_{\rm \rho \ge 10^6 g cm^{-3}} \rho_* |{q}_{\rm neu}|\alpha  d^3 x,
 \end{equation}
 respectively. Here, $\epsilon$ is the specific internal energy. 
 \begin{figure}[t]
 	\includegraphics[scale=0.7]{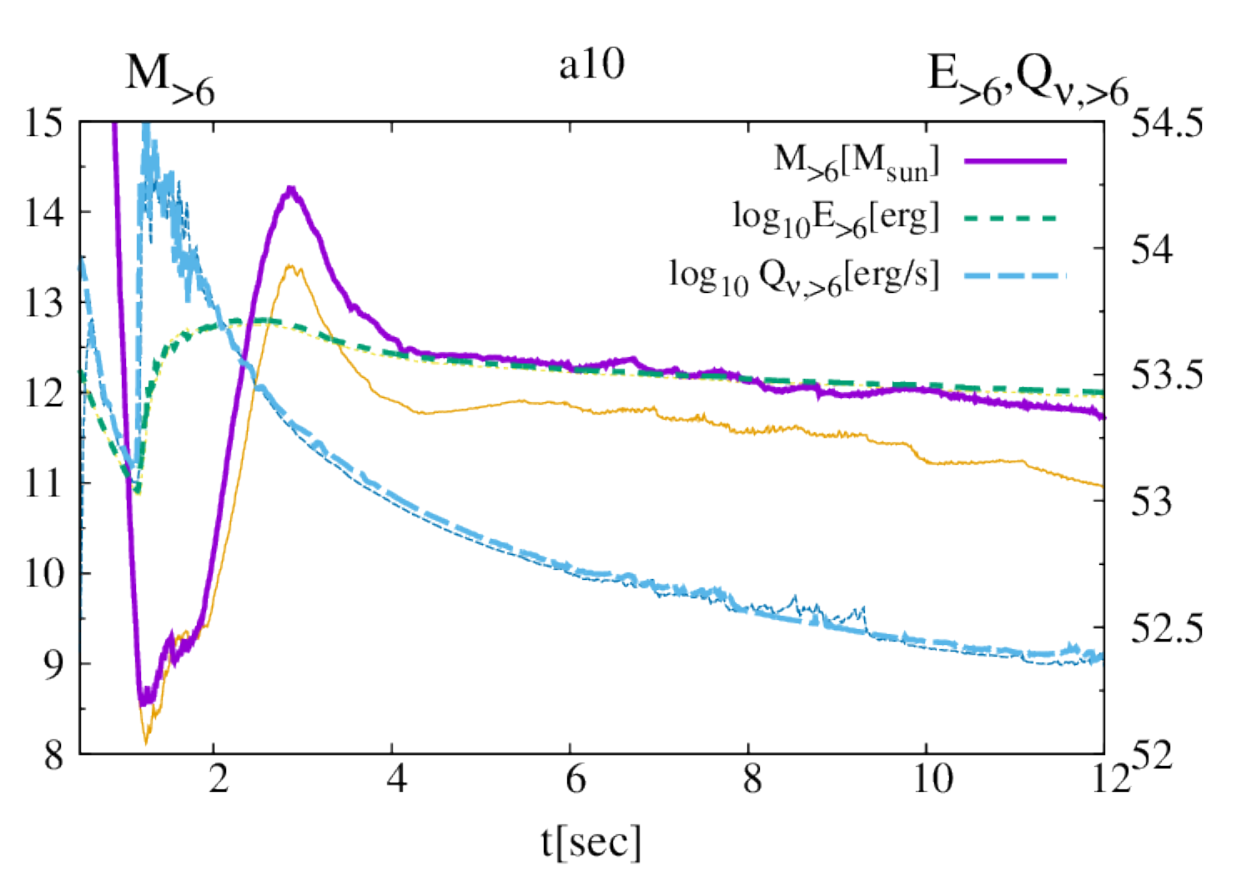}
 	\caption{Time evolution of several quantities for model a10. The solid, dotted and dashed curves show the total mass, internal energy and neutrino emission rate of the region with $\rho \geq 10^6~{\rm g/cm^3}$, respectively. The thin-solid, thin-dotted and thin-dashed curves show these quantities for the low-resolution case, respectively.}
 	\label{fig:tevo10}
 \end{figure}

\begin{figure*}[htbp]
	\begin{minipage}{0.3\hsize}
		\begin{center}
%			\hspace*{-5em}  
			\includegraphics[width=1\linewidth,angle=0,trim=60 20 60 0]{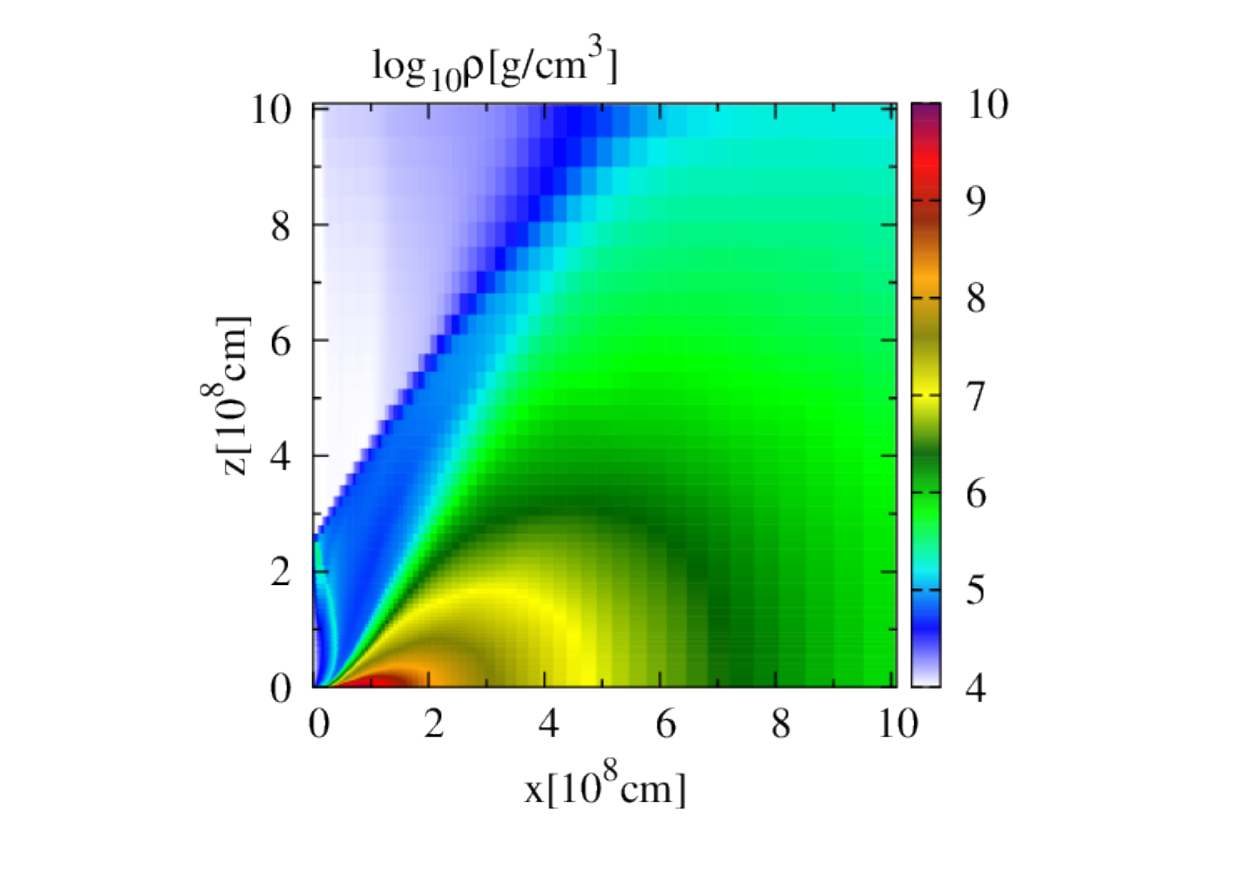}
		\end{center}
	\end{minipage}
	\begin{minipage}{0.3\hsize}
		\begin{center}
%			\hspace*{-5em} 
			\includegraphics[width=1\linewidth,angle=0,trim=60 20 60 0]{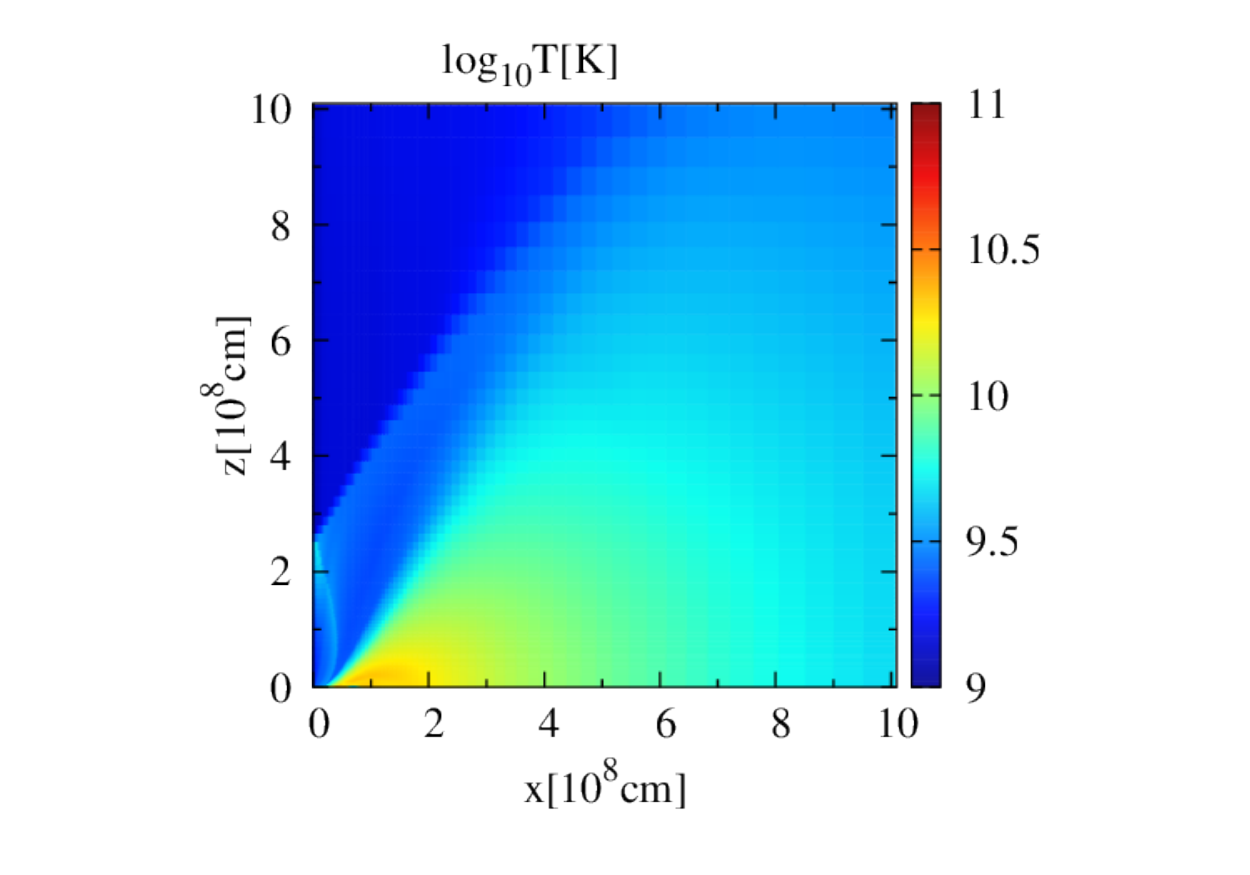}
		\end{center}
	\end{minipage}
	\begin{minipage}{0.3\hsize}
		\begin{center}
%			\hspace*{-5em} 
			\includegraphics[width=1\linewidth,angle=0,trim=60 20 60 0]{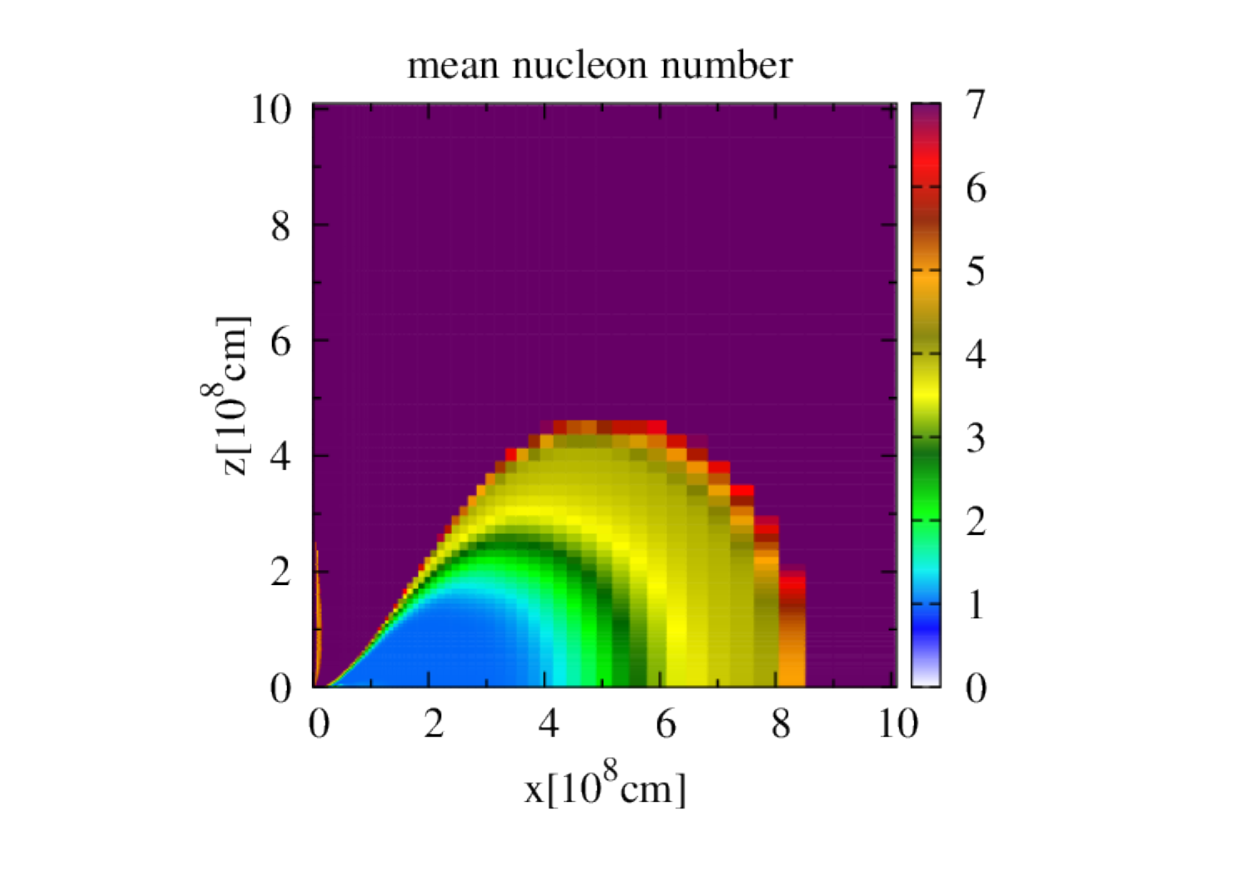}
		\end{center}
	\end{minipage}
	\caption{The rest-mass density (left), temperature (middle) and mean nucleon number (right) profiles of the torus for model a10 at $t=10$ s.}
	\label{fig:torus}
\end{figure*}
 Figure \ref{fig:tevo10} shows the time evolution of $M_{>6}$ (solid curve), $E_{>6}$ (dotted curve) and $Q_{\nu,>6}$ (dashed curve) for model a10. The bubble starts expanding and collides with the torus at $t\approx 1$ s.
 Then, the torus is heated up by the shock. 
 For $t\gtrsim 4$ s, the shock front passes through the torus.  
Thereafter, the torus is gradually cooled by the neutrino emission. The mass of the torus is $\sim 12M_\odot$ (approximately $8\%$ of the initial rest mass of the CO core) and neutrino cooling timescale $\tau_{\nu} \equiv E_{>6}/Q_{\nu,>6} \approx 10$ s at $t=10$ s. {At the end of the simulation,} the torus relaxes to a quasi-stationary state.

Figures \ref{fig:torus} displays the distributions of the density (left), temperature (middle) and mean nucleon number (right) of the torus for model a10 at $t=10$ s.
The maximum density and temperature of the torus are  $6.7 \times 10^{9}~{\rm g/cm^3}$ and $2.1\times 10^{10}$ K, respectively. Due to such high temperature, nearly all the region of the torus is in a state of NSE.
The distribution of the mean nucleon number shows that the torus is composed primarily of protons, neutrons and $^4$He generated by photodissociation. Total masses of nucleons (protons and neutrons) and $^4$He in the torus at $t=10$ s are $\sim 9.4M_\odot$ and $2.8M_\odot$, respectively.

\subsubsection{model a07}
  \begin{figure*}[htbp]
 	\begin{minipage}{0.3\hsize}
 		\begin{center}
% 			\hspace*{-5em}  
 			\includegraphics[width=1\linewidth,bb=0 0 576 403,angle=0,trim=120 20 120 60]{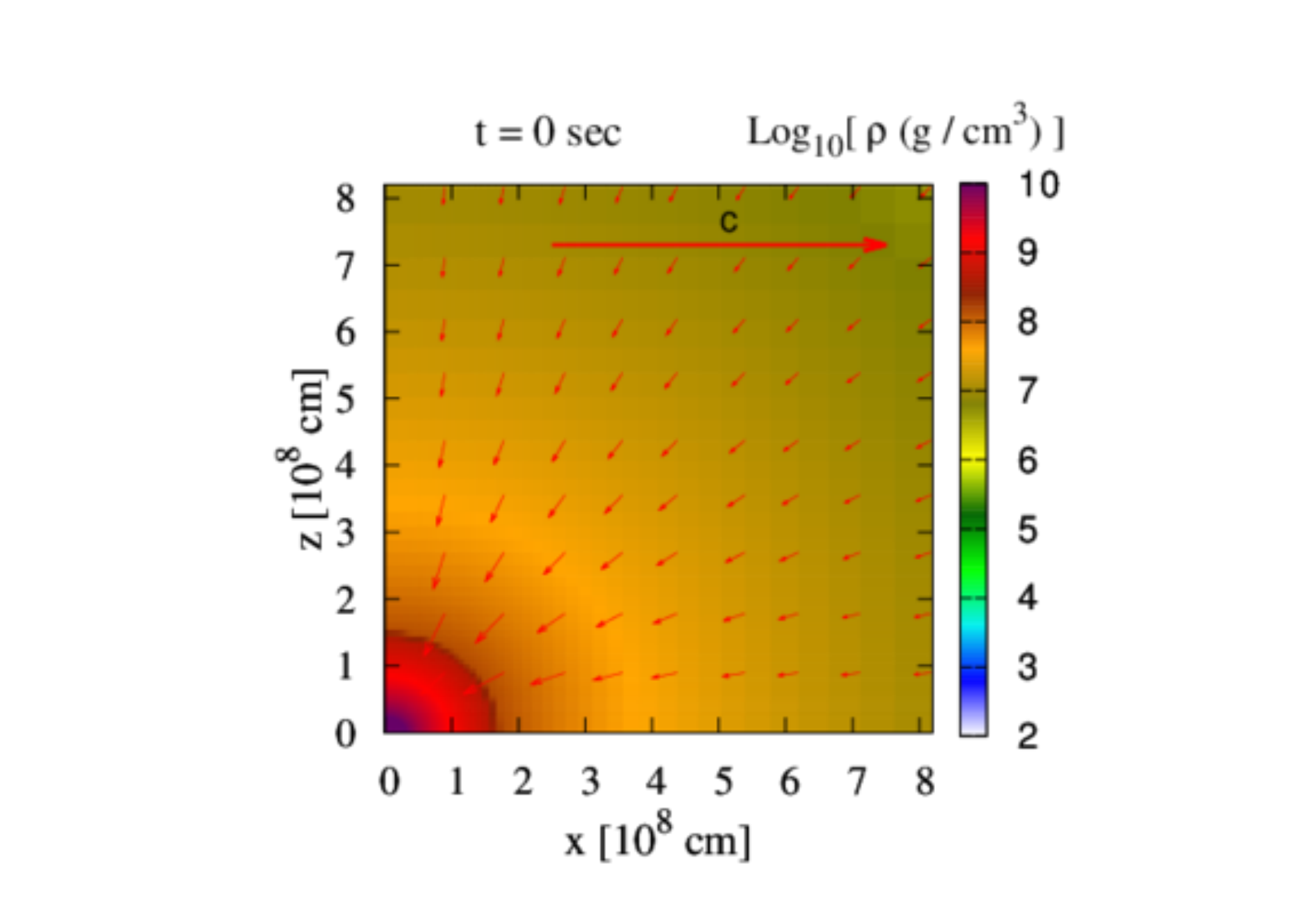}
 		\end{center}
 	\end{minipage}
 	\begin{minipage}{0.3\hsize}
 		\begin{center}
% 			\hspace*{-5em} 
 			\includegraphics[width=1\linewidth,bb=0 0 576 403,angle=0,trim=120 20 120 60]{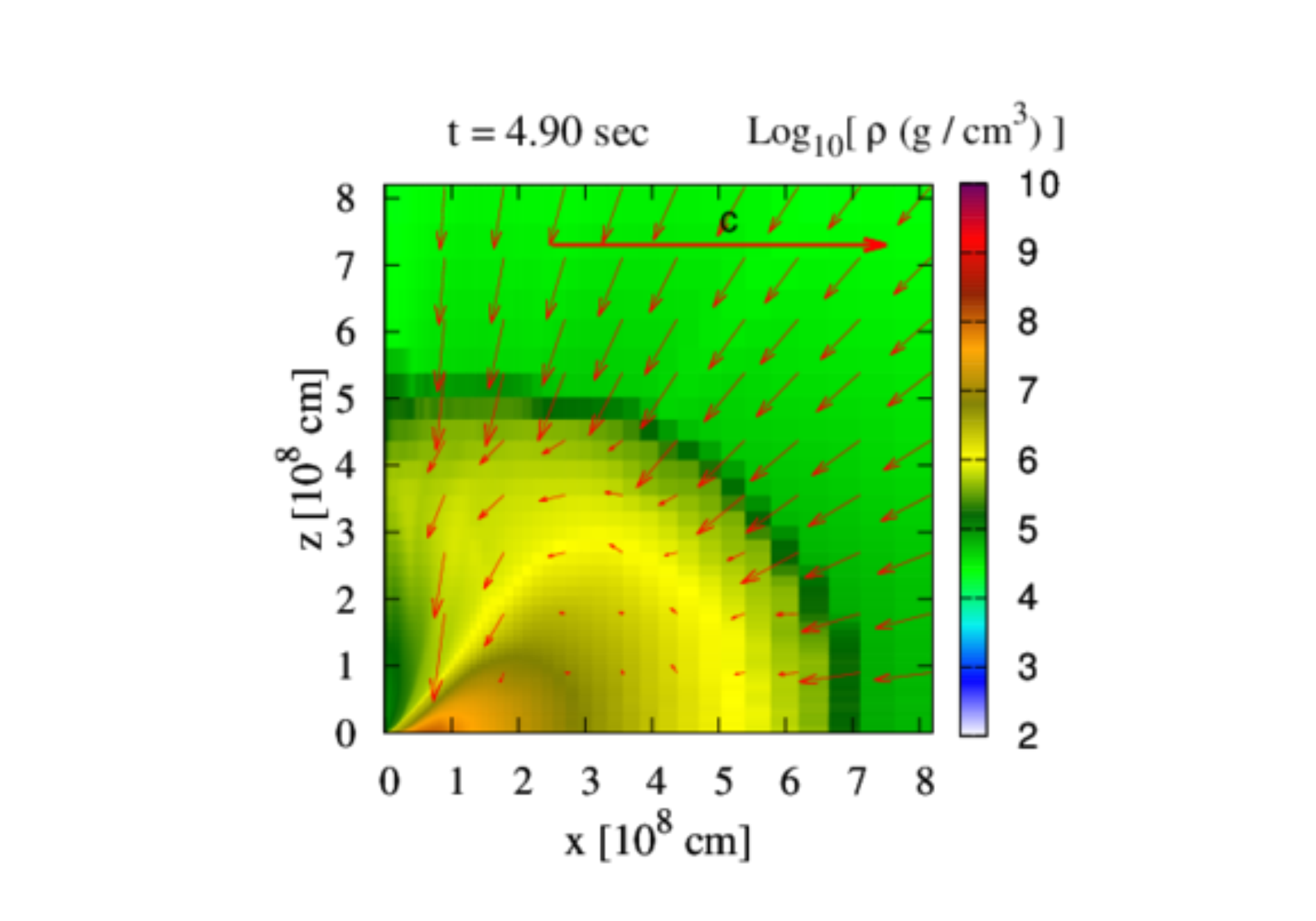}
 		\end{center}
 	\end{minipage}
 	\begin{minipage}{0.3\hsize}
 		\begin{center}
% 			\hspace*{-5em} 
 			\includegraphics[width=1\linewidth,bb=0 0 576 403,angle=0,trim=120 20 120 60]{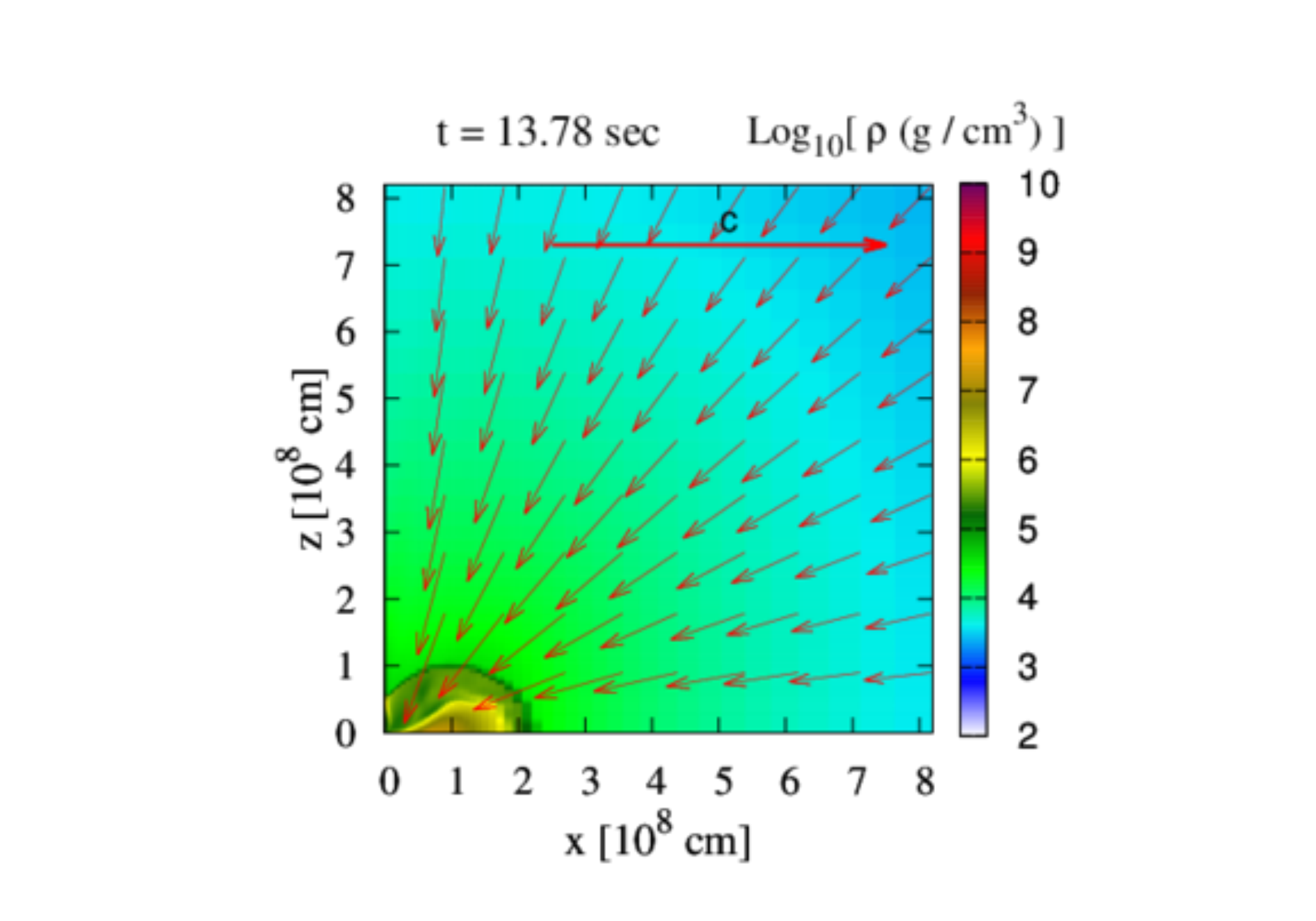}
 		\end{center}
 	\end{minipage}
 	\caption{Same as Figure~\ref{fig:collapse} but for model a07.}
 	\label{fig:collapse_o7}
 \end{figure*}
\begin{figure*}[htpb]
	\begin{minipage}{0.3\hsize}
		\begin{center}
%			\hspace*{-5em}  
			\includegraphics[width=1\linewidth,angle=0]{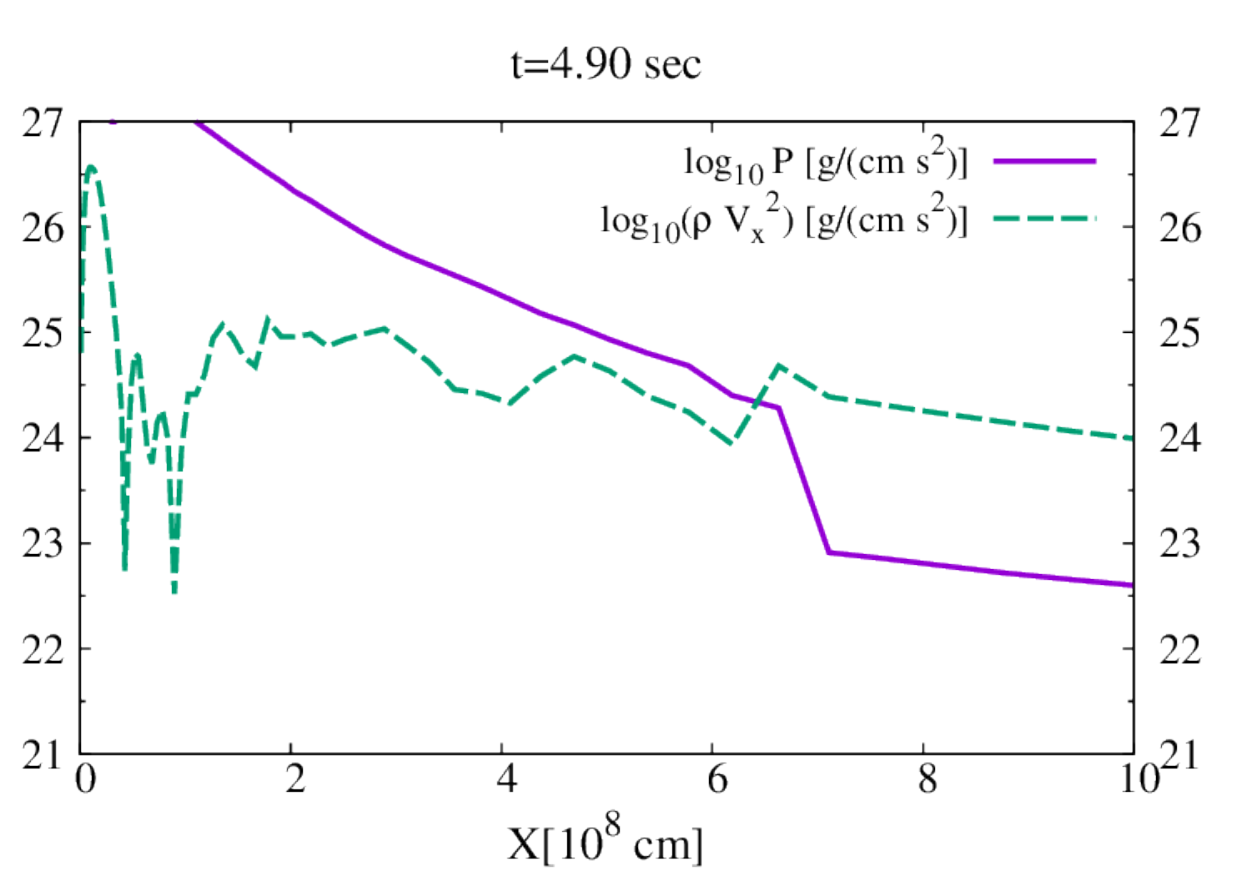}
		\end{center}
	\end{minipage}
	\begin{minipage}{0.3\hsize}
		\begin{center}
%			\hspace*{-5em} 
			\includegraphics[width=1\linewidth,angle=0]{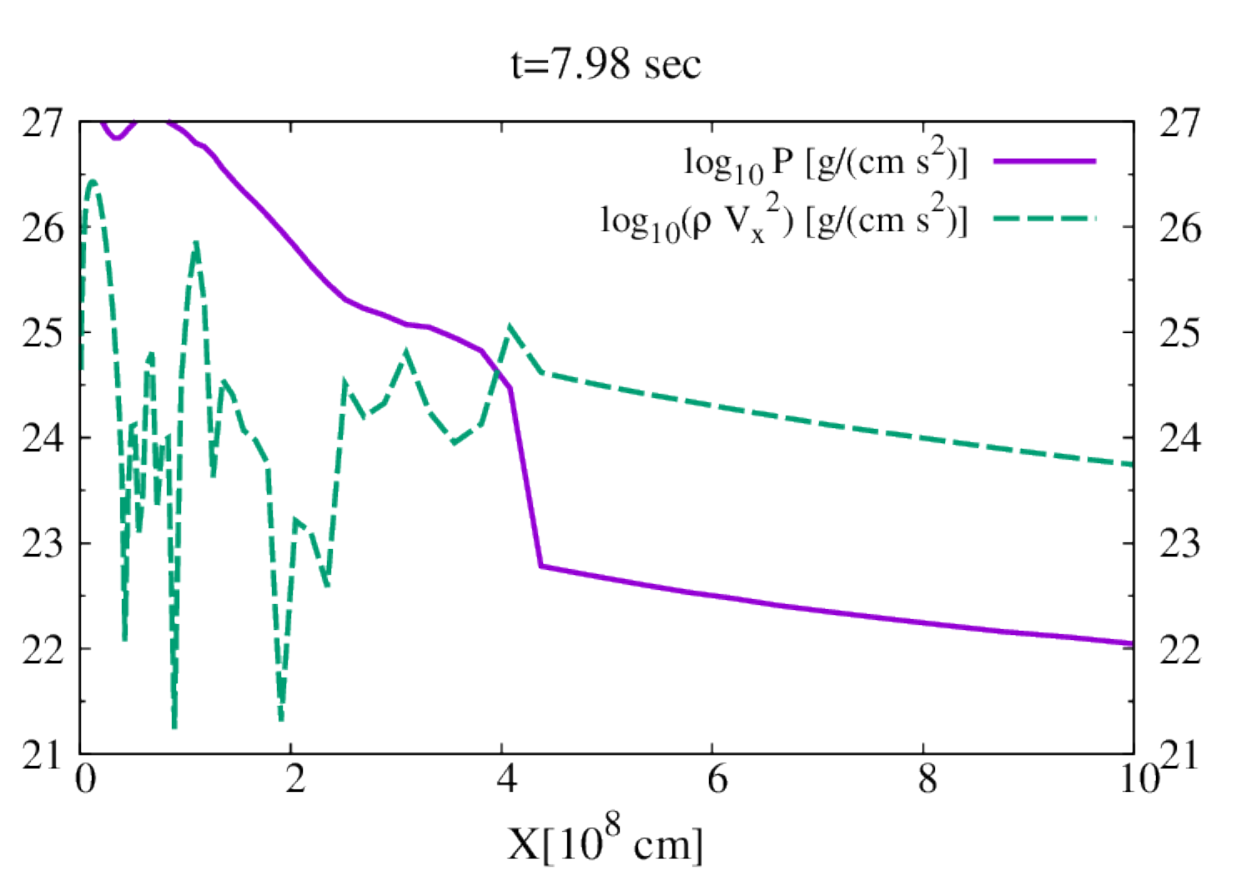}
		\end{center}
	\end{minipage}
	\begin{minipage}{0.3\hsize}
		\begin{center}
%			\hspace*{-5em} 
			\includegraphics[width=1\linewidth,angle=0]{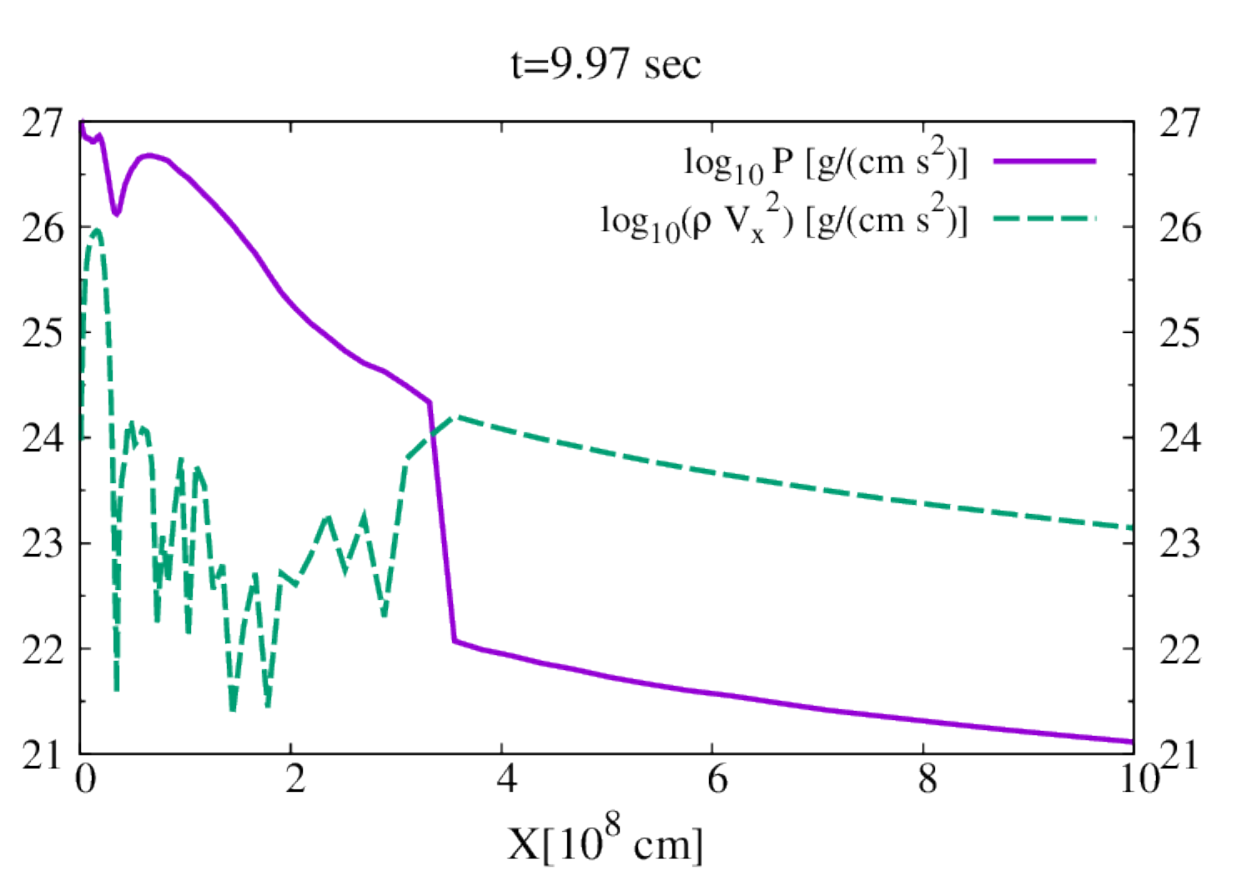}
		\end{center}
	\end{minipage}
	\caption{Snapshots of the pressure (solid curve) and ram pressure ($\rho V_{X}^2$, dashed curve) profiles in the equatorial plane during the shock propagation for model a07. Here, $V_{X} \equiv u^X/u^t$.}
	\label{fig:tpres_o07}
\end{figure*}
 \begin{figure}[htbp]
 	\includegraphics[scale=0.7]{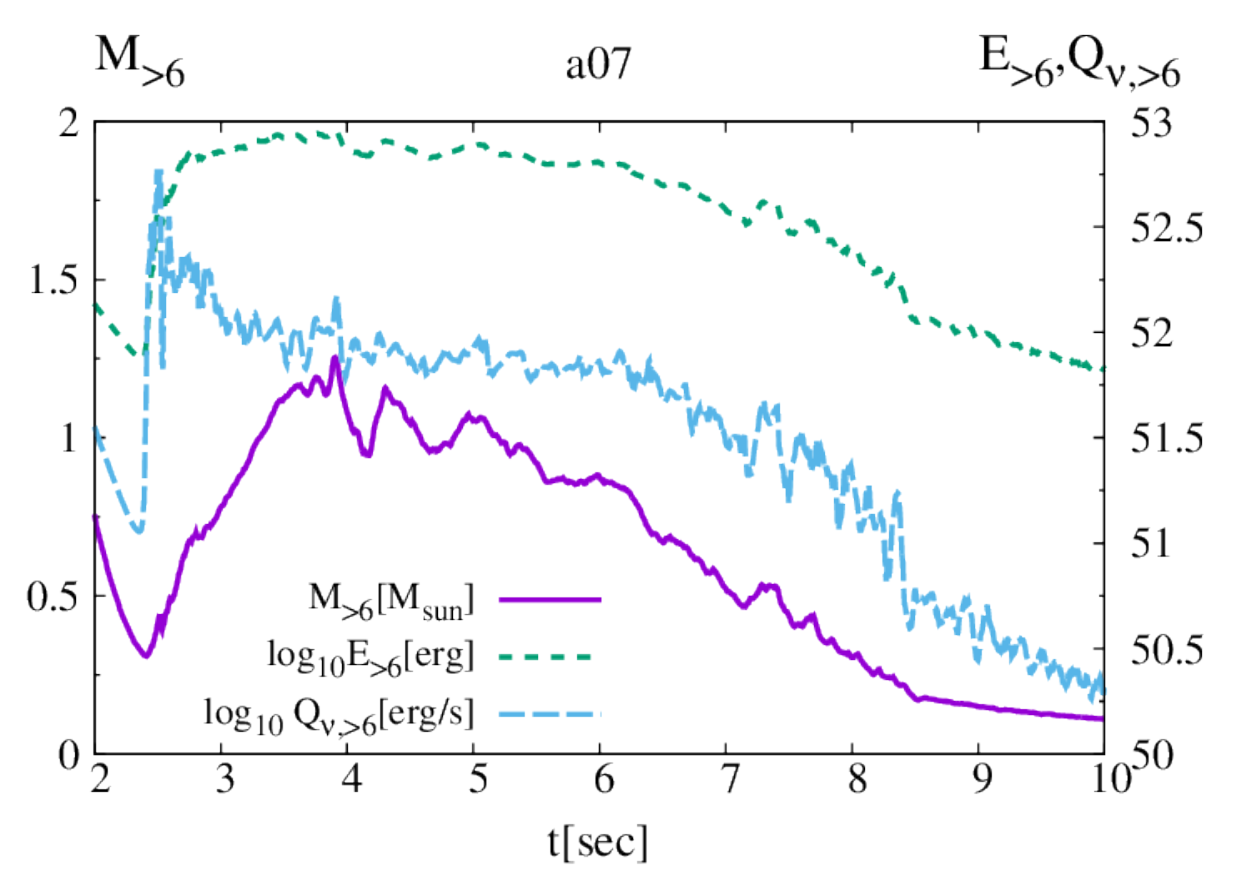}
 	\caption{Time evolution of several quantities of the torus for model a07.  The solid, dotted and dashed curves show the total mass, internal energy and neutrino emission rate of the region with $\rho \geq 10^6~{\rm g/cm^3}$, respectively.}
 	\label{fig:tevo07}
 \end{figure}
 Figure \ref{fig:collapse_o7} displays the snapshots of the density profile for model a07.
 For model a07, although the torus and outflow are formed in the same process as model a10, the shock front stagnates at a radius of $< 10^9$ cm in several seconds after the BH formation and then falls by the ram pressure of the infalling matter at $t\gtrsim 8$ s. 
 {Figure~\ref{fig:tpres_o07} illustrates that the pressure behind the shock is defeated by the ram pressure, and hence, the shock front falls toward the BH.} 
 
Figure~\ref{fig:tevo07} shows that the torus evolves approximately in the same process as model a10 for $t < 8$ s. 
However, for $t>8$ s, the shock front falls down and the torus material accretes onto the BH. 
As a result, a small torus with mass $\sim 0.1M_\odot$ is formed at $t=10$ s for this model.

\subsubsection{model a04} 
 \begin{figure*}[htbp]
 	\begin{minipage}{0.3\hsize}
 		\begin{center}
 %			\hspace*{-5em}  
 			\includegraphics[width=1\linewidth,bb=0 0 576 403,angle=0,trim=120 20 120 60]{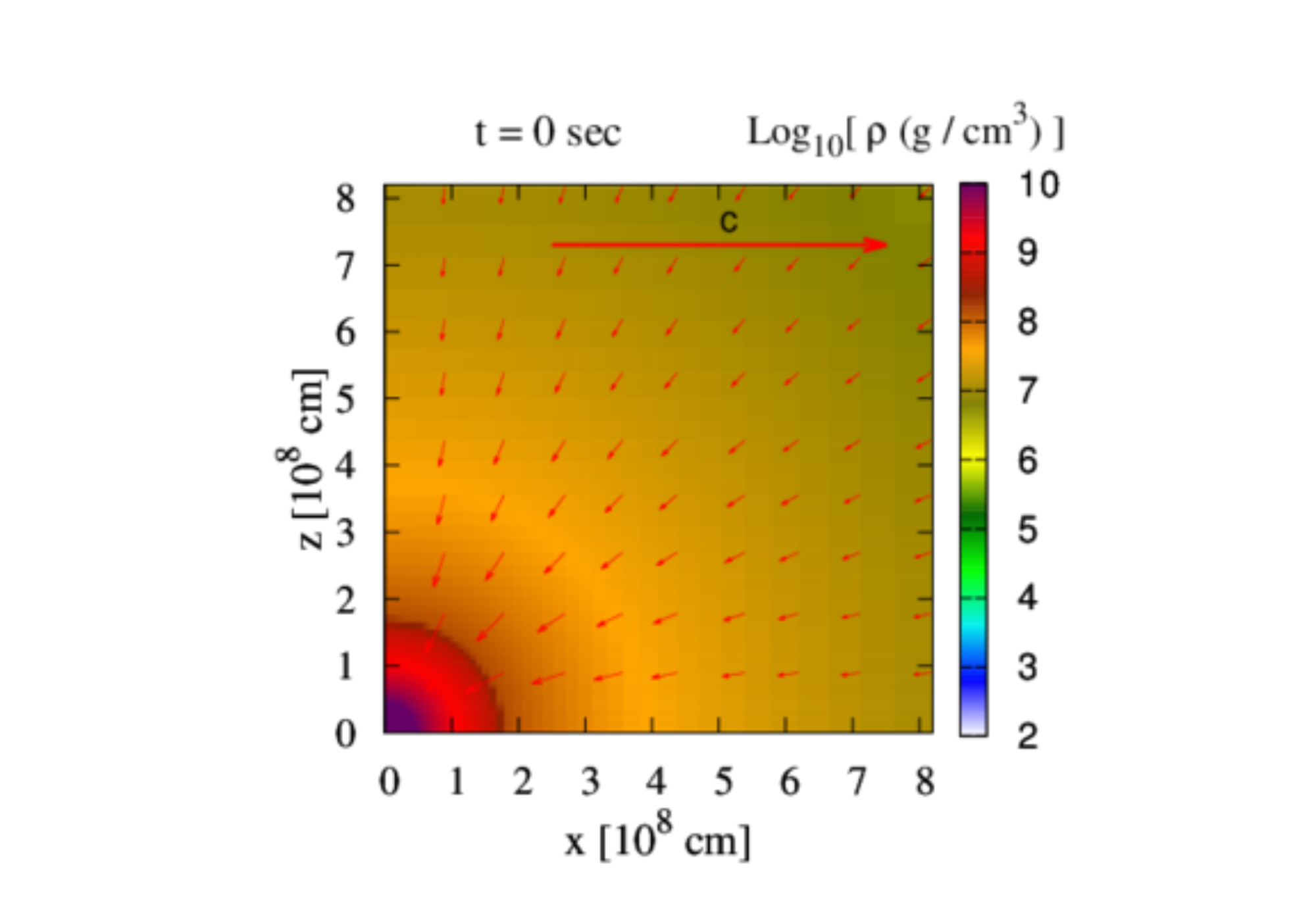}
 		\end{center}
 	\end{minipage}
 	\begin{minipage}{0.3\hsize}
 		\begin{center}
% 			\hspace*{-5em} 
 			\includegraphics[width=1\linewidth,bb=0 0 576 403,angle=0,trim=120 20 120 60]{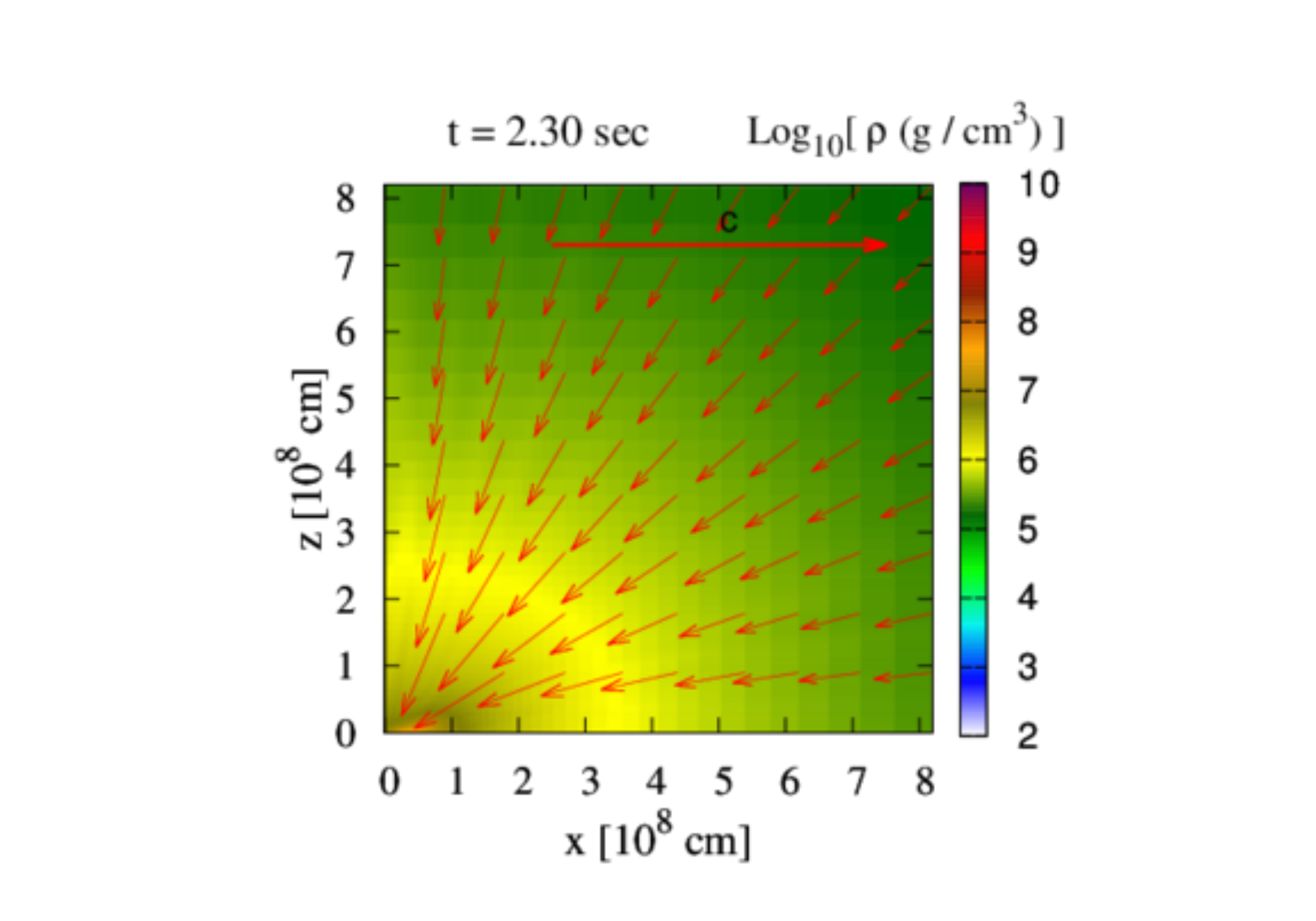}
 		\end{center}
 	\end{minipage}
 	\begin{minipage}{0.3\hsize}
 		\begin{center}
% 			\hspace*{-5em} 
 			\includegraphics[width=1\linewidth,bb=0 0 576 403,angle=0,trim=120 20 120 60]{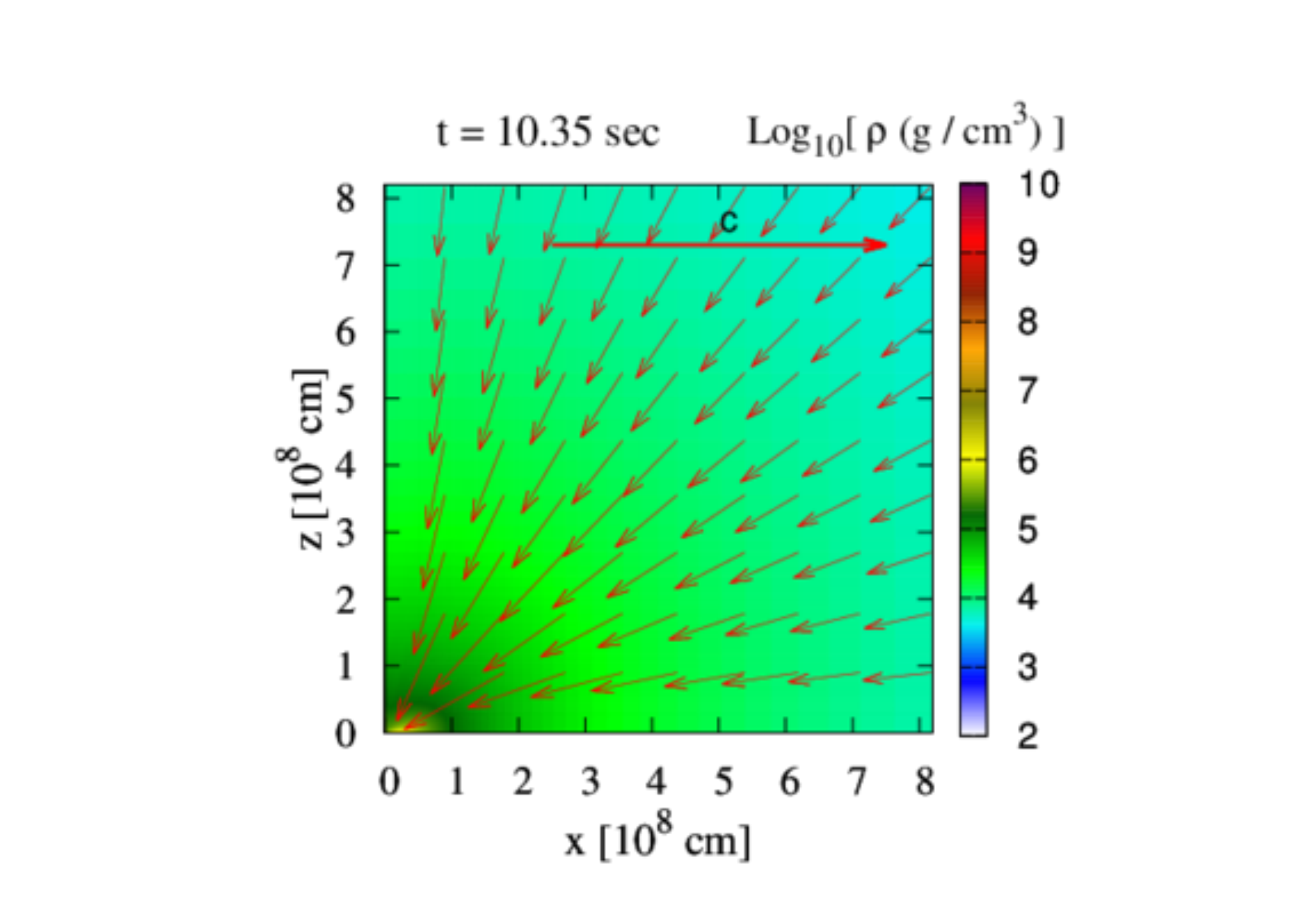}
 		\end{center}
 	\end{minipage}
 	\caption{{Same as Figure~\ref{fig:collapse_o7}, but for model a04.}}
 	\label{fig:collapse_o4}
 \end{figure*}
\begin{figure}[htbp]
	\includegraphics[scale=0.7]{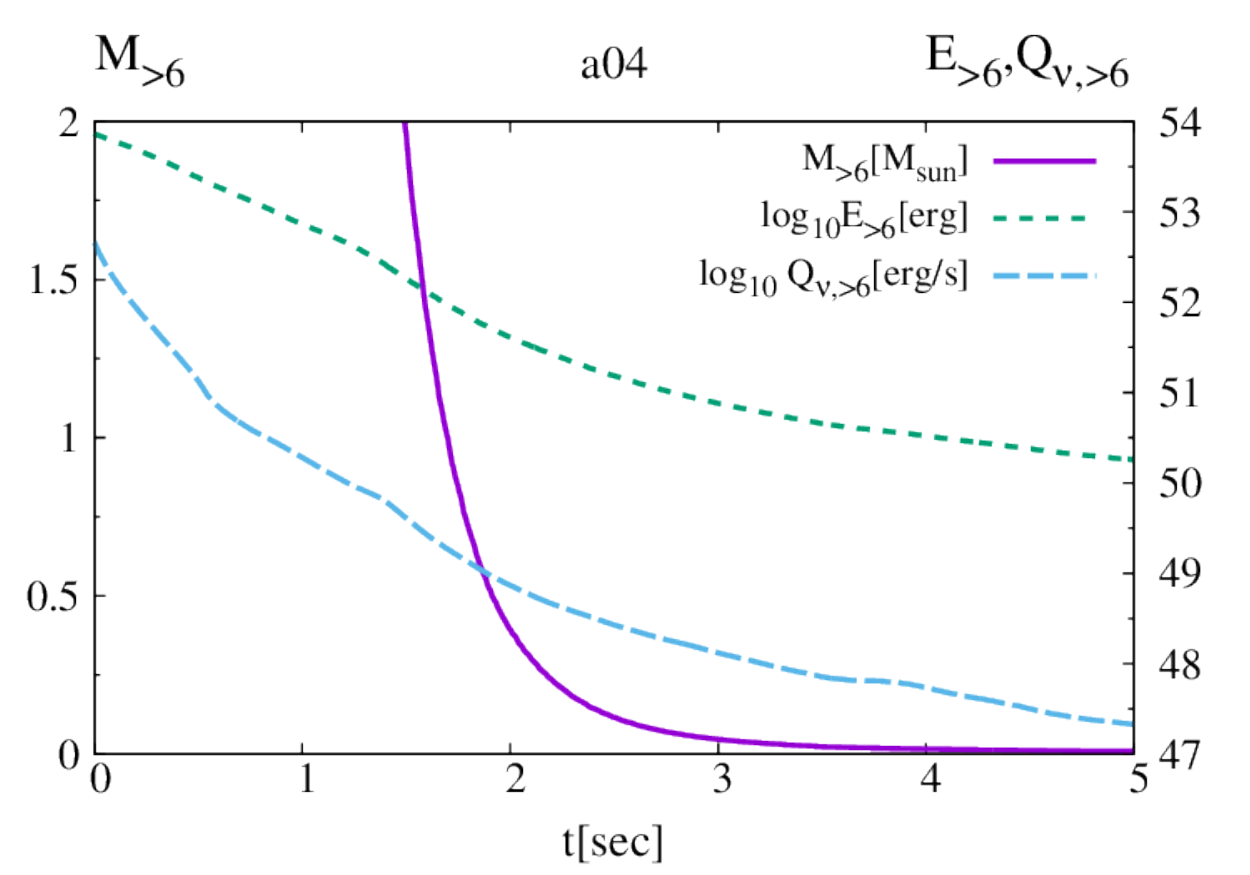}
	\caption{{Same as Figure~\ref{fig:tevo07}, but for model a04.}}
	\label{fig:tevo04}
\end{figure}
As we illustrate in Figures~\ref{fig:collapse_o4} and \ref{fig:tevo04}, no outflow occurs and no torus is formed for model a04. This is due to the fact that the specific angular momentum of the fluid elements is too small to remain around the BH. This result is consistent with what we expected in Section~\ref{models}.

In conclusion, {for the case that we consider only the effects of the hydrodynamical instabilities on the transport of angular momentum} in the stellar evolution calculation for rapidly rotating VMSs, approximately $8\%$ of the initial rest mass of the CO core remains as the torus surrounding the BH and strong expanding shocks are formed. However, if the {additional mechanisms (e.g., TS dynamo)} decreases the rotation velocity even by $30\%$, finally a torus with an appreciable amount of mass is not formed and outflow does not occur or accretes immediately onto the BH even if it is driven at the formation of a torus.
\subsection{Properties of the BH}
\label{BH}
\begin{table}[htpb]
	\caption{The key parameters of the BH for all the models at $t=10$ s. 
		$M_{\rm BH}$ and  $q_{\rm BH}$ are the mass and dimensionless spin parameter of the BH, respectively. Here, $M_{\rm BH}^{\rm e}$ and  $q_{\rm BH}^{\rm e}$ are the values estimated in Section~\ref{models}.}
		\begin{center}
	\begin{tabular}{|c||c|c||c|c|} \hline
		Model & $M_{\rm BH}$ & $q_{\rm BH}$ & $M_{\rm BH}^{\rm e}$& $q_{\rm BH}^{\rm e}$\\ \hline \hline
		a10 & $131M_{\odot}$ & 0.79 & $122M_{\odot}$ & 0.83 \\ \
		a07 & $148M_{\odot}$ & 0.76 & $147M_{\odot}$ & 0.72\\ \
		a04 & $148M_{\odot}$ & 0.44 & $150M_{\odot}$ & 0.44\\  \hline
	\end{tabular}
	\end{center}
	\label{tab:tor}
\end{table}
We list the quantities of the BH for all the models at $t=10$ s in Table \ref{tab:tor}. 
Here, $M_{\rm BH}^{\rm e}$ and  $q_{\rm BH}^{\rm e}$ are the values estimated in Section~\ref{models}. 
{For models a10 and a07, the mass of the BH is larger than the value estimated in Section~\ref{models}. 
This is likely due to the fact that each fluid element of the VMS core falls toward the BH with an elliptical orbit, and then, some fluid elements fall into the BH even if their specific angular momentum is larger than $j_{\rm ISCO}$. The pressure and geometry of the torus may also affect the result.}

{For $q_{\rm BH}$, model a07 has larger value than $q_{\rm BH}^{\rm e}$. It is natural because the fluid elements with large specific angular momentum, which we did not expect to fall into the BH, actually fall and if the whole core collapses to form a BH, $q_{\rm BH}$ is estimated to become $\sim 0.78~(>q_{\rm BH}^{\rm e})$ for model a07 (see Section~\ref{models}). However,  for model a10, $q_{\rm BH}$ is lower than $q_{\rm BH}^{\rm e}$. This is due to the neutrino emission from the torus as  we describe in the following. }

{For model a10, a large amount of neutrinos are emitted from the torus at the time when the torus is heated up by the shock (see Figure~\ref{fig:tevo10}). 
Since the torus material is rotating with a relativistic velocity ($0.3$--$0.4~c$), the neutrinos take away a part of the angular momentum of the torus due to the relativistic beaming effect.
Thus, the specific angular momentum of the fluid elements which accretes onto the BH decreases. On the other hand, for model a07, the neutrino emission is not as strong as for model a10 because the mass of the torus is much  smaller than for model a10 (see Figure~\ref{fig:tevo07}). Thus, the specific angular momentum of the fluid elements is not significantly taken away by neutrinos. 
For model a04, all the fluid elements of the core accerete to the BH with little neutrino emission, and thus, the BH with an expected spin is formed.}
\subsection{Convergence}
We remark the convergence property of the numerical results. 
The thin-solid, thin-dotted and thin-dashed curves in Figure~\ref{fig:tevo10} show the time evolution of $M_{>6}$, $E_{>6}$ and $Q_{\nu,>6}$ for the low-resolution case for model a10, respectively. 
{It is found that these quantities depend only weakly on the grid resolution.} 
We check that the differences between the different resolution models at $t=10$\,s for $M_{>6}$, $E_{>6}$ and $Q_{\nu,>6}$ are within $6.0\%$, $4.3\%$ and $5.6\%$, respectively. 
We also check that the difference of $M_{\rm BH}$ and $q_{\rm BH}$ between the different resolution models at $t=10$ s are within $0.4\%$ and $1.3\%$, respectively. 
Thus, the agreement with different resolution models is reasonably achieved for these quantities.

\section{discussion}
\label{discussion}
\subsection{Evolution of the torus}
\label{torus}
As we found in the previous section, after the gravitational collapse, a massive torus is formed around the BH and  expanding shocks are launched for model a10. First, we briefly discuss the possible evolution scenario of the torus due to the viscous effect. 

{The last two panels of Figure \ref{fig:collapse} show that the infall velocity of the fluid elements inside the shock is much smaller than that of the outside. This suggests that the torus does not feel strong ram pressure of the infalling matter, and thus, we expect that the evolution of the torus would be similar to that of an isolated torus.}

 We consider the possible viscous effect associated with magnetohydrodynamics (MHD) turbulence in the torus  \citep[e.g.,][]{2013ApJ...772..102H,2014ApJ...784..121S,2016MNRAS.456.2273S,2016MNRAS.457..857S}, which is not taken into account in our present study but it could play an important role for the evolution of the torus surrounding the BH in reality. We evaluate the strength of the viscosity by Shakra $\&$ Suniyaev's alpha viscosity model \citep{1973A&A....24..337S}. In this model, the shear viscous coefficient can be written as
\begin{equation}
\nu = \alpha_{\rm vis} c_{s} H,
\end{equation}
where $\alpha_{\rm vis}$ is the so-called $\alpha$-viscosity parameter, $c_s$ and $H$ are 
the sound speed and the vertical scale height of the torus, respectively.

Assuming that the evolution of the torus could be described by a standard accretion disk theory \citep{1973A&A....24..337S}, 
the viscous timescale is estimated by
\begin{eqnarray}
t_{\rm vis} \sim \alpha_{\rm vis}^{-1} \left (   \frac{R_{\rm torus}^3}{GM_{\rm BH} }  \right )^{1/2}\left (   \frac{H}{R_{\rm torus}}  \right )^{-2} \nonumber \\
\approx 19~{\rm s}~\left (   \frac{ \alpha_{\rm vis}  }{ 0.01  }  \right )^{-1} \left (   \frac{ M_{\rm BH}  }{ 130M_\odot  }  \right )^{-1/2} \nonumber \\
\times \left (   \frac{ R_{\rm torus}  }{2\times 10^8~{\rm cm}  }  \right )^{3/2}\left (   \frac{H/R_{\rm torus} }{  1/3 }  \right )^{-2},
\label{vis}
\end{eqnarray}
where $M_{\rm BH}$ and $R_{\rm torus}$ are the mass of the BH and the typical radial scale of the torus,  respectively. 
Equation~(\ref{vis}) shows that the viscous accretion timescale is of the order of 10 s for a plausible value of $\alpha_{\rm vis} =O(0.01)$ for model a10. Here, $\alpha_{\rm vis}=O(0.01)$ is suggested to be a typical value for accretion disks by a number of high-resolution MHD simulations \citep{2013ApJ...772..102H,2014ApJ...784..121S,2016MNRAS.456.2273S,2016MNRAS.457..857S}.

The mass accretion could occur in this timescale and matter in the outer part of the torus receives the angular momentum from its inner part. Then, the torus expands and a part of its mass would be ejected.  \cite{2013MNRAS.435..502F} and \cite{2015MNRAS.448..541J} showed that a fraction ($\sim 20\%$) of the matter of a torus surrounding a BH could be ejected by the viscous effect in the absence of strong cooling effects. Also, recent general-relativistic MHD simulations showed that the total amount of the ejected mass could reach $20$--$40\%$ of the initial torus mass \citep{2017PhRvL.119w1102S,2018arXiv180800461F}.

{We expect that in the expanding torus and ejecta, the recombination of light elements would occur and abundant thermal energy would be released \citep[e.g.,][]{2013MNRAS.435..502F}.} % and the mass ejection would become stronger.
The thermal energy released for the recombination of protons and neutrons to $^4$He is
\begin{equation}
E_{{\rm 2p+2n \rightarrow ^{4}He}} \approx 1.4 \times 10^{52} \left (   \frac{ M_{\rm p+n } }{M_\odot   }  \right ) {\rm erg},
\end{equation}
where $M_{\rm p+n}$ is the total mass of reacted protons and neutrons which is assumed to be $\sim 10\%$ of the torus mass.

Hence, if the nuclear reaction of protons and neutrons to $^4$He  occurs even with a few $M_\odot$, 
explosive energy as the same order as explosion energy of hypernovae ($\sim 10^{52}$ erg) is released. 
If such an amount of thermal energy is efficiently deposited, {the mass ejection would become stronger, and thus, the ejecta would inject more energy to the envelope of the progenitor star. Then, the luminosity of the explosion would be high enough to be observed (see Section \ref{explum}).} Since we found that the cooling timescale by the neutrino emission of the torus is $\gtrsim 10$ s, long-term simulations including both the neutrino cooling and viscous heating are necessary to follow the detailed time evolution of the torus. This is an interesting topic to be explored in the future.

{We note that recent MHD simulations of the gravitational collapse of magnetized massive Population III stars and SMSs suggest that the magnetic-field effects could launch a jet  \citep{2007PASJ...59..771S,2017PhRvD..96d3006S,2018arXiv180707970S}. Such jets may also significantly inject the energy in the envelope. }

\subsection{Explosion luminosity}
\label{explum}
When the outflow and ejecta collide with the envelope,  shocks are formed and propagate through the envelope.  
We have found in this study that the energy of the outflow is much less than the possible injection energy from the torus ejecta.
Thus, in the following, we take into account the energy injection only from the torus.

After the shock propagation, the matter in the  envelope behind the shock is heated up, 
and when the shock reaches the surface of the envelope, it could be observed as a SN explosion. 
Since the expanding envelope is hydrogen-rich, this process is similar to the Type IIP SNe. 
Here, we estimate the bolometric light curve of the hypothetical explosion by using the method of  \cite{1980ApJ...237..541A} and \cite{1993ApJ...414..712P} and following Appendix 4 of \cite{2013ApJ...778...67N}. 

\cite{1980ApJ...237..541A} and \cite{1993ApJ...414..712P} analytically formulated the light curve model of the spherically expanding shock-heated ejecta. 
This formulation assumes that the explosion energy injected by the ejecta, $E_{\rm exp}$, is equally distributed into kinetic and internal energy, i.e., $E_{\rm exp}/2=E_{\rm int}=E_{\rm kin}$ where $E_{\rm int}$ and $E_{\rm kin}$ are the total internal energy and kinetic energy of the envelope, respectively.  
The radius of the surface of the envelope, $R(t)$ is written as
\begin{equation}
R(t) \equiv R_0 + v_{\rm sc} t,
\end{equation}
where $R_0$ is the initial radius of the envelope (for our model, $R_0\approx 3.6\times 10^{14}~{\rm cm}$) and $v_{\rm sc}$ is the expansion velocity. 

Assuming the uniform density profile and the homologous expansion, we have
\begin{eqnarray}
\rho(t) &=& \frac{3M_{\rm env}}{4 \pi R_0^3} \left (   \frac{ R(t)  }{ R_0  }  \right )^{-3} , \\
v(r,t)&=& \frac{r}{R(t)}v_{\rm sc},
\end{eqnarray}
where $r$ and $M_{\rm env}$ are the radial coordinate and total mass of the envelope, respectively. By using the equation of the kinetic energy, $E_{\rm kin} = \int_{0}^{R(t)} \rho v^2/2~dV$, $v_{\rm sc}$ is obtained  as
\begin{eqnarray}
v_{\rm sc} = \sqrt{\frac{10E_{\rm kin}}{3M_{\rm env}}} &\approx& 2.4 \times 10^8 ~{\rm cm/s}\nonumber \\ 
 &\times& \left (   \frac{E_{\rm exp}   }{10^{52}~{\rm erg}   }  \right )^{\frac{1}{2}}\left (   \frac{M_{\rm env}   }{ 150M_\odot  }  \right )^{-\frac{1}{2}}.
\label{lum-1}
\end{eqnarray}

From the first law of thermodynamics, the thermal evolution of the envelope could be described by 
\begin{equation}
 \frac{\partial{  e } }{\partial{   t } }  +P \frac{\partial{ } }{\partial{   t } } \left (   \frac{   1}{\rho   }  \right ) = - \frac{\partial{  L } }{\partial{   m_r } },
 \label{th1}
\end{equation}
where $e,~P,~L$ and $m_r$ are the specific internal energy, pressure, luminosity and enclosed mass, respectively.  We suppose that the envelope is radiation-pressure-dominant, i.e., $\rho e=3P=aT^4$ where $a$ is the radiation constant. 
In the diffusion approximation, $L$ is written as
\begin{equation}
L=-\frac{4\pi r^2 a c}{3\kappa \rho}  \frac{\partial{ T^4 } }{\partial{  r } }.
\end{equation}
Here, $\kappa$ is the opacity. 
For simplicity, we assume that $\kappa$ is written as 
\begin{eqnarray}
  \kappa=\left\{ \begin{array}{ll}
   \kappa_{\rm T} \approx 0.27~{\rm cm^2~g^{-1}} \,\, &(T\geq T_{\rm ion}),\\
   0  &(T<T_{\rm ion}),\\
\end{array} \right.
   \label{kap}
\end{eqnarray} 
where $\kappa_{\rm T}$ is the Thomson scattering opacity of the envelope (we assumed that the mass fractions of $^1$H and $^4$He of the envelope are $\sim 0.34$ and $\sim0.63$, respectively) and $T_{\rm ion}$ is the hydrogen recombination temperature. We take $T_{\rm ion}=6000$ K and for the region $T<T_{\rm ion}$, we assume that the recombination of hydrogen atoms occurs and this region becomes optically transparent \citep{1980NYASA.336..335W,1985SvAL...11..145L}.

We define the expansion time, $t_{\rm e}$, and the photon diffusion time, $t_{\rm d}$, as
\begin{eqnarray}
t_{\rm e} &\equiv& \frac{R_0}{v_{\rm sc}} \approx 1.7 \times 10^6 ~{\rm s} \label{texp}\\ 
 &\times& \left (   \frac{ R_0  }{ 4\times 10^{14}~{\rm cm}  }  \right )\left (   \frac{  E_{\rm exp} }{ 10^{52}~ {\rm erg} }  \right )^{-\frac{1}{2}}\left (   \frac{ M_{\rm env}  }{ 150M_\odot  }  \right )^\frac{1}{2}, \nonumber \\ 
t_{\rm d} &\equiv& \sqrt{\frac{9\kappa M_{\rm env}}{2\pi^3 v_{\rm sc} c}} \label{tdiff} \\
&\approx& 4.0 \times 10^{7} {~\rm s} \left (   \frac{  E_{\rm exp} }{10^{52}~ {\rm erg}   }  \right )^{-\frac{1}{4}}\left (   \frac{ M_{\rm env}  }{150M_\odot   }  \right )^{\frac{3}{4}}. \nonumber
\end{eqnarray}
Then, from Equations (\ref{th1}) and (\ref{kap}), the time evolution of the photosphere radius, $R_{\rm ph}(t)$,  is obtained as
\begin{eqnarray}
R_{\rm ph}(t)=\left\{ \begin{array}{ll}
R(t) \,\, &(t\leq t_{\rm i}),\\
R_{\rm ion}(t)  &(t>t_{\rm i}),
\label{rph}
\end{array} \right.
\end{eqnarray}
where
\begin{equation}
R_{\rm ion}(t)^2 = v_{\rm sc}^2 \left \{ \tau_{\rm i} \tau \left ( 1+  \frac{\tau_{\rm i}^2   }{ 3t_{\rm d}^2 }  \right ) - \frac{\tau^4}{3t_{\rm d}^2} \right \}.
\end{equation}
Here, $\tau \equiv t+t_{\rm e}$ and $\tau_{\rm i}\equiv t_{\rm i}+t_{\rm e}$. $t_{\rm i}$ is the time at whith the surface effective temperature of the photosphere drops to $T_{\rm ion}$. 
The method for calculating $t_{\rm i}$ is described in the next paragraph.

The time evolution of the bolometric luminosity is obtained as
\begin{eqnarray}
L(t)=\left\{ \begin{array}{ll}
L_0{\rm exp} \{ -(t^2+2t_{\rm e}t )/t_{\rm d}^2\} \,\, &(t\leq t_{\rm i}),\\
4 \pi R_{\rm ion}(t)^2 \sigma_{\rm SB} T_{\rm ion}^4   &(t>t_{\rm i}),
\label{lum}
\end{array} \right.
\end{eqnarray}
where
\begin{eqnarray}
L_0 &=& \frac{t_{\rm e}E_{\rm exp}}{t_{\rm d}^2} \approx 1.0\times 10^{43}~{\rm erg/s} \nonumber \\
&\times& \left (   \frac{ R_0  }{ 4\times 10^{14}~{\rm cm}  }  \right )\left (   \frac{  E_{\rm exp} }{ 10^{52} ~{\rm erg} }  \right ) \left (   \frac{ M_{\rm env}  }{ 150M_\odot  }  \right )^{-1}.
\end{eqnarray}
Here, $\sigma_{\rm SB}$ is the Stefan Boltzmann constant. The effective temperature of the photosphere, $T_{\rm eff}$, can be given by the relation $L=4\pi R_{\rm ph}^2 \sigma_{\rm SB} T_{\rm eff}^4$ and 
$t_{\rm i}$ can be determined by solving the condition $T_{\rm eff}(t_{\rm i}) =T_{\rm ion}$. 
For the case of $t_{\rm i}^2\ll t_{\rm d}^2$ and $t_{\rm e}t_{\rm i} \ll t_{\rm d}^2$, $t_{\rm i}$ is approximately given by
\begin{eqnarray}
t_{\rm i}+t_{\rm e}  &\sim& \left (   \frac{ L_0 }{ 4\pi \sigma_{\rm SB} v_{\rm sc }^2}  \right )^\frac{1}{2} T_{\rm ion}^{-2} \\
&\approx& 1.4 \times 10^7~{\rm s} \left (   \frac{ R_0  }{ 4\times 10^{14}~{\rm cm}  }  \right )^\frac{1}{2}. \nonumber
\end{eqnarray}
 \begin{figure}[t]
	\includegraphics[scale=0.7]{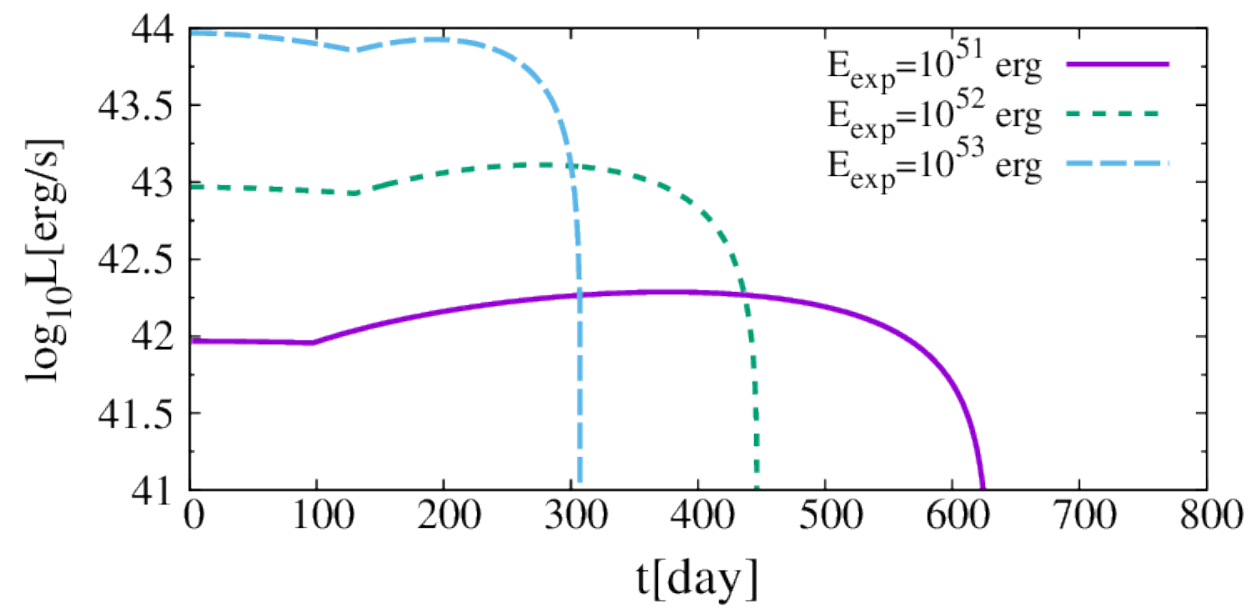}
	\caption{Time evolution of the bolometric luminosity described in Equation (\ref{lum}). We take $R_0 =3.6\times 10^{14}$ cm and $M_{\rm env}=150M_\odot$. The solid, dotted and dashed curves show the light curves for the case of of $E_{\rm exp}=10^{51},~10^{52}$ and $10^{53}$ erg, respectively. }
	\label{fig:lc}
\end{figure}
$R_{\rm ion}$ takes a maximum at $t=t_{\rm max}$ where
\begin{eqnarray}
t_{\rm max}&+&t_{\rm e} \sim \left (   \frac{ 3 }{4 }  \right )^{1/3} \tau_{\rm i}^{1/3} t_{\rm d}^{2/3}  \\
&\approx& 2.5 \times 10^7 ~{\rm s} \nonumber \\
&\times& \left (   \frac{ R_0  }{ 4\times 10^{14}~{\rm cm}  }  \right )^\frac{1}{6} \left (   \frac{  E_{\rm exp} }{ 10^{52}~ {\rm erg} }  \right )^{-\frac{1}{6}} \left (   \frac{ M_{\rm env}  }{ 150M_\odot  }  \right )^{\frac{1}{2}}. \nonumber 
\end{eqnarray}

$R_{\rm ph}$ at $t=t_{\rm max}$ is given by
\begin{eqnarray}
R_{\rm ph}(&t_{\rm max}&) \sim \left (   \frac{ 3  }{ 4 }  \right )^{2/3}v_{\rm sc} \tau_{\rm i}^{2/3} t_{\rm d}^{1/3} \\
&\approx& 3.9\times 10^{15}~{\rm cm} \nonumber \\ 
&\times& \left (   \frac{ R_0  }{ 4\times 10^{14}~{\rm cm}  }  \right )^\frac{1}{3}\left (   \frac{  E_{\rm exp} }{ 10^{52}~ {\rm erg} }  \right )^{\frac{5}{12}} \left (   \frac{ M_{\rm env}  }{ 150M_\odot  }  \right )^{-\frac{1}{4}}, \nonumber
\end{eqnarray}
where we neglect the term $\tau_{\rm i}^2/3t_{\rm d}^2 ~(\ll 1)$.
At this time, the bolometric luminosity also takes a maximum value,
\begin{eqnarray}
L_{\rm max} &\sim& 4 \pi R_{\rm ph}(t_{\rm max})^2 \sigma_{\rm SB} T_{\rm ion}^4 \\
&\approx& 1.4 \times 10^{43} ~{\rm erg/s} \nonumber \\ 
&\times& \left (   \frac{ R_0  }{ 4\times 10^{14}~{\rm cm}  }  \right )^\frac{2}{3} \left (   \frac{  E_{\rm exp} }{ 10^{52} ~{\rm erg} }  \right )^{\frac{5}{6}} \left (   \frac{ M_{\rm env}  }{ 150M_\odot  }  \right )^{-\frac{1}{2}}.  \nonumber
\end{eqnarray}

Figure \ref{fig:lc} shows the time evolution of the bolometric luminosity described in Equation (\ref{lum}). This shows that for the case of $E_{\rm exp}=10^{52}$ erg, the bolometric luminosity becomes $\approx 10^{43}$ erg/s and its duration is of order yrs. These luminosity and timescale are similar to those of the peculiar hydrogen-rich SN \citep{2017Natur.551..210A}.

\subsection{Ringdown gravitational waves associated with the formation of BHs}
\label{GW}
Our previous studies for the gravitational collapse of rapidly rotating SMSs showed that burst gravitational waves are emitted in the formation process of rotating BHs and the total radiated energy, $\Delta E$, is of the order of $10^{-6}M_{\rm BH}c^2$ (\cite{PhysRevD.94.021501,2017PhRvD..96h3016U}, see also \cite{2017PhRvD..96d3006S,2018arXiv180707970S}). 
Our preliminary simulations show that during the BH formation in the collapse of rapidly rotating VMSs, gravitational waves are also emitted in a similar manner to those of SMSs. 

Here, we briefly estimate the frequency and amplitude of gravitational waves and discuss its observability for model a10. The frequency of gravitational waves, $f_{\rm gw}$, can be approximated by the frequency of the axisymmetric mode of the ringdown oscillation associated with the formed BH \citep{2009CQGra..26p3001B}.  Inserting $M_{\rm BH}= 131M_\odot$ and $a_{\rm BH}=0.79M_{\rm BH}$ to the Equation (97) of \cite{2009CQGra..26p3001B}, we obtain $f_{\rm gw}\approx 99$ Hz. 

Following \cite{PhysRevD.94.021501}, we suppose that the gravitational wave amplitude would be approximately written as 
\begin{eqnarray}
h&\sim& \frac{4GM_{\rm BH}}{c^2 D}\sqrt{\epsilon_{\rm gw}}, \\
 &\approx&  1.6\times 10^{-22} \left (   \frac{M_{\rm BH}}{131M_\odot}  \right ) \left (\frac{D }{ 100~{\rm Mpc}  }  \right )^{-1} \left (   \frac{  \epsilon_{\rm gw} }{ 10^{-6}  }  \right )^{1/2},\nonumber
\end{eqnarray}
where $D$ is the luminosity distance to the source and $\epsilon_{\rm gw}\equiv \Delta E /M_{\rm BH}c^2$. 
We take $D=100$ Mpc, which is the same order as the luminosity distance of the peculiar hydrogen-rich SN \citep{2017Natur.551..210A}. Thus, if $\epsilon_{\rm gw}=O(10^{-6})$ and the collapse of a VMS occurs at $D \sim 100$ Mpc, gravitational waves observed will be in the sensitive observation band of ground-based gravitational wave detectors (e.g., Advanced LIGO; \cite{2017CQGra..34d4001A}), in particular future ground-based detectors such as Einstein telescope \citep{2011CQGra..28i4013H}. It is our future work to investigate the observability of these gravitational waves in more detail.

\subsection{Remark}
\label{GRB}
{It is uncertain whether rapidly rotating VMSs in isolation could result in rapidly rotating CO cores such as those investigated in this work. 
	On one hand, the slow rotation rates of neutron stars \citep{2000ApJ...528..368H},  white dwarfs \citep{2008A&A...481L..87S} and red giant cores \citep{2012A&A...548A..10M} suggest that
	the angular momentum transport in a star is more efficient than that expected from the hydrodynamical instabilities. 
	On the other hand, progenitor stars of GRBs are considered to be massive stars which have a rapidly rotating core  \citep{1993ApJ...405..273W,1999ApJ...524..262M}. 
	That is, even though there are efficient angular momentum transport mechanisms, special evolution paths which can form a rapidly rotating core should exist (e.g., binary merger \citep{1998ApJ...502L...9F,2005ApJ...623..302F} and  chemically homogeneous evolution  \citep{2005A&A...435..967Y,2006ApJ...637..914W}). 
	Thus, VMSs with rapidly rotating CO cores are likely to be formed by such special evolution scenarios.}

\section{conclusion}
\label{conclusion}
We explored the gravitational collapse of rotating VMSs in axisymmetric numerical relativity.
We selected a progenitor ZAMS star with the initial mass of $M_{\rm ZAMS} =320M_\odot$ and rotating rigidly with the rotation velocity of $50\%$ of the Kepler rotation at its surface.　

1D stellar evolution calculation is performed from the ZAMS stage until the central temperature reaches ${\rm log}_{10} T_{\rm c}~[{\rm K}]\approx 9.2$ including the effects of {angular momentum transport induced by  hydrodynamical instabilities}. At this stage, we mapped the resulting 1D stellar evolution models onto 2D grids of axisymmetric gravitational collapse simulations as the initial conditions.
In the stellar evolution stage, {we neglect the effects of other additional  angular momentum transport mechanisms.} 

{The additional mechanisms may increase the efficiency of angular momentum transport and decrease the final core angular velocity (for example, it is suggested that the TS dynamo would decrease it approximately by one order of magnitude \citep{2018ApJ...857..111T}). To consider the cases for which  the angular velocity is decreased due to such effects,} we simulated two additional models for which the angular velocity is {multiplied by a factor of $0.7$ (model a07) and $0.4$ (model a04) in addition to the original model (model a10)}.

 We found that for all the models, a BH is formed promptly after the gravitational collapse. For model a10, a fraction of the accreted matter forms a torus surrounding the remnant BH and drives an outflow. {The outflow expands to form shocks in the core.}
In a few seconds after the BH formation, the torus relaxes to a quasi-stationary state with mass $\approx 12M_\odot$  composed of protons, neutrons and $^4$He generated by photodissociation. On the other hand, for models a07 and a04, finally only a small or no torus is formed and outflow does not occur or accretes immediately onto the BH even if it is driven at the formation of a torus. This is because the specific angular momentum of all the fluid elements is too small to form a large torus and strong outflow.

{We analyzed the parameters of the remnant BH formed after the gravitational collapse.
	For models a10 and a07, the resulting mass of the BH is larger than {the expectation based on a method  in~\cite{0004-637X-818-2-157}}. 
	{This is because in our expectation, we regard each fluid element of the VMS core as a test particle, and assume that it has a circular orbit around the hypothetically formed BH.
	However, in reality, each fluid element of the VMS core falls in an elliptical orbit to the central BH and feels the pressure force. Hence, a fraction of the additional fluid elements fall into the BH even though they have specific angular momentum larger than the value of ISCO. 
	For the spin of the BH, model a07 has a larger value than the expectation. It is natural because the fluid elements with large specific angular momentum, which we do not expect to fall into the BH, actually fall.
	On the other hand, model a10 has a lower value than the expectation. 
	This is due to the neutrino emission from the torus.}
	Since the torus rotating with relativistic velocity ($0.3$--$0.4~c$) emits a large amount of neutrinos for model a10, a part of the angular momentum of the fluid elements accreting to the BH is taken away by the neutrinos due to the relativistic beaming effect. Thus, the resulting spin of the BH for model a10 becomes a lower value. On the other hand, for model a07, since the effect of the neutrino emission is minor, the angular momentum of the fluid elements is not significantly taken away by neutrinos. 
	For model a04, all the fluid elements of the core accrete to the BH with little neutrino emission, and hence, a  BH with expected mass and spin is formed.}

We discussed a possible evolution process of the torus for model a10. 
Because a strong outflow is driven soon after the the BH formation in this model, the torus does not feel strong ram pressure of the infalling matter, and thus, the evolution of the torus would be similar to that of an isolated torus.  
Assuming the $\alpha$-viscosity model and that the evolution of the torus could be described by a standard accretion disk theory, we found that the viscous accretion timescale is of the order of 10 s for a plausible value of $\alpha_{\rm vis} =O(0.01)$ for model a10. 
Because the neutrino cooling timescale is longer than the viscous timescale, the torus is likely to expand and a part of its mass is ejected in the viscous timescale. 
We expect that in the expanding torus and ejecta, the recombination of light elements would occur and abundant thermal energy would be released. If the nuclear reaction of protons and neutrons {to $^4$He} occurs in the ejecta even with a few $M_\odot$, explosive energy as the same order as explosion energy of hypernovae ($\sim 10^{52}$ erg) is released. {If such an amount of thermal energy is efficiently deposited, the mass ejection would become powerful, and thus, the ejecta would inject significant energy to the envelope.}

{We estimated bolometric luminosity and the timescale supposing that the energy injection from the torus to the envelope occurs and after the energy injection, the envelope would be heated up and start expanding like in a SN explosion. We found that if the ejecta injects energy of the order of $10^{52}$ erg, the bolometric luminosity and timescale are of the order of $10^{43}$ erg/s and  yrs, respectively. }

We discussed the possibility for observing gravitational waves associated with the BH formation. We estimate the frequency and amplitude of gravitational waves for model a10 and find that 
if the total radiated energy is $O(10^{-6}) M_{\rm BH} c^2$ and the collapse of a VMS takes at $\sim 100$ Mpc, the frequency and amplitude are $\sim 100$ Hz and $\sim 10^{-22}$. 
These values are in the sensitive observation band of ground-based gravitational wave detectors, in particular, future ground-based detectors such as Einstein telescope. 
Exploring the observability of these gravitational waves in more detail is our future work.

{\em Acknowledgments}
: We are grateful to Yudai Suwa for a helpful discussion. Numerical computations were performed on the supercomputer XC50 at CfCA of NAOJ, and XC40 at YITP of Kyoto University. 
This work was supported by Grant-in-Aid for Scientific Research~(Grants No.~16H02183~16K17706~16H05341~15H00782) of Japanese MEXT/JSPS. KT was supported by the Japan Society for the Promotion of Science (JSPS) Overseas Research Fellowships.

%%%%%%%%%%%%%%%%%%%% REFERENCES %%%%%%%%%%%%%%%%%%
%\bibliography{biblio}
\bibliography{myrefs}

\appendix
\section{Numerical Test : PISN simulation}
\label{A}

\begin{figure*}[htpb]
	\begin{minipage}{0.5\hsize}
		\begin{center}
			\hspace*{-5em}  
			\includegraphics[angle=0,scale=0.7]{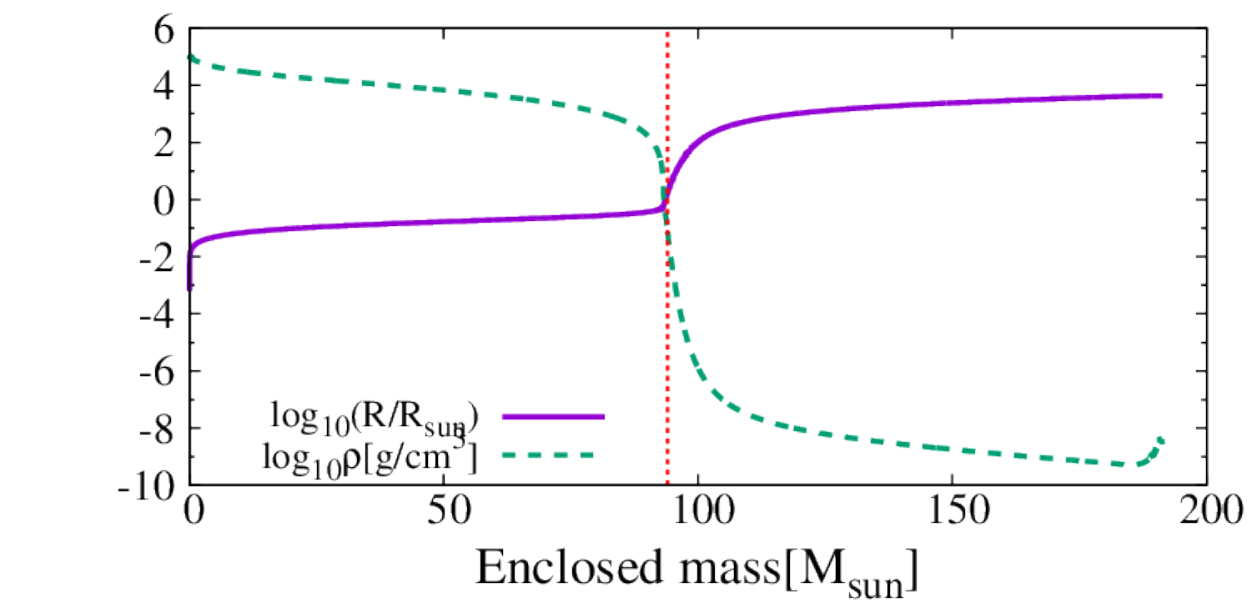}
		\end{center}
	\end{minipage}
	\begin{minipage}{0.5\hsize}
		\begin{center}
			\hspace*{-5em} 
			\includegraphics[angle=0,scale=0.7]{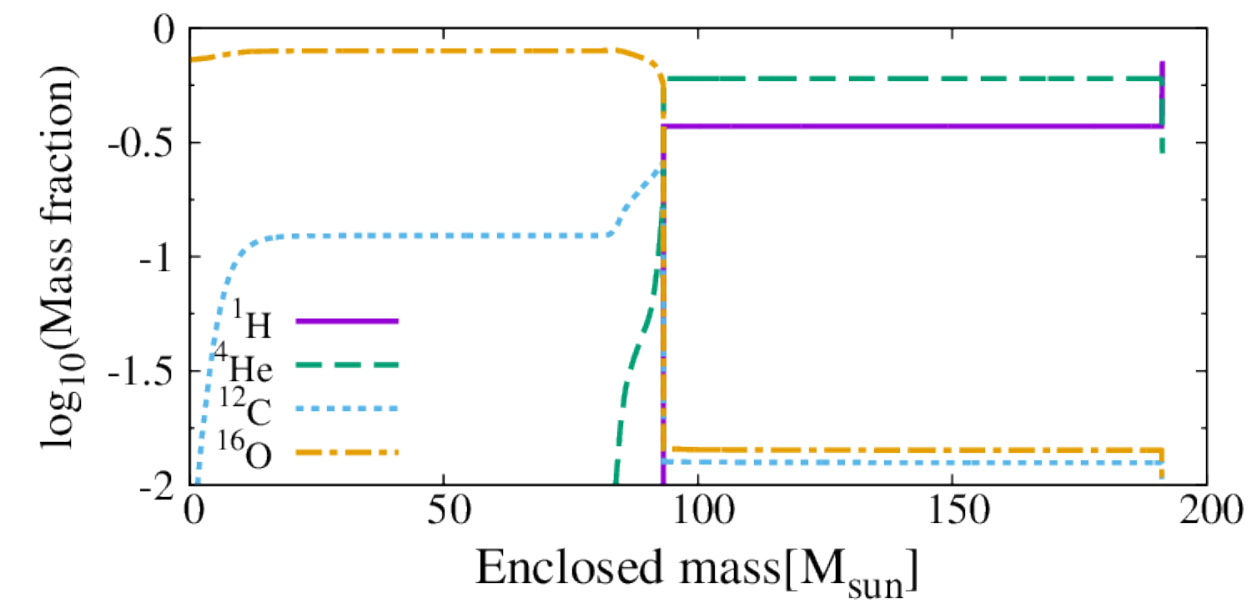}
		\end{center}
	\end{minipage}
	\caption{Progenitor profile of the model for the test calculation as functions of the enclosed mass at the end of stellar evolution calculation. Left panel shows the density (solid curve) and radius (dashed curve) distribution, respectively. Right panel shows the chemical distribution. The solid, dashed, dotted and dashed-dotted curves show the chemical abundance of $^1$H, $^4$He, $^{12}$C and $^{16}$O, respectively. The vertical red dotted line of the left panel denotes $X_{\rm max}$ of the test calculation. }
	\label{fig:piprogen}
\end{figure*}

For a test calculation of our 2D gravitational collapse code, we performed a PISN simulation of a non-rotating VMS with the initial mass of $M_{\rm ZAMS} =200M_\odot$. For the initial rotation profile of the progenitor ZAMS star, we employed the rigid rotation with the rotation velocity of $30\%$ of the Kepler rotation at its surface (angular velocity $\approx 1.7\times 10^{-4}~{\rm s}^{-1}$).
Figure~\ref{fig:piprogen} shows that the density, radius profile (left panel) and chemical distribution (right panel) of the progenitor star at the end of stellar evolution calculation. The total mass of the star at this stage is $M \approx 190M_\odot ~(R_M\approx 2.8\times 10^7~{\rm cm})$ and the outer edge of the CO core is located at  $\approx 3.9\times 10^{10}~\rm{cm}~(\approx 1400R_M)$.
 
We performed the 1D-spherical calculation in addition to the 2D calculation and compare the results of each calculation. For the 1D calculation, the 1D-spherical general-relativistic Lagrangian hydrodynamic code \citep{1997ApJ...475..720Y} with 47 isotopes reaction network are used. 
For the 2D calculation, the grid parameters for the axisymmetric numerical simulation are set to be $(\Delta X_0,~\eta,~X_{\rm in},~X_{\rm max}) = (2R_M,~1.017,~40R_M,~3400R_M)$.
The vertical red dotted line of the left panel of Figure~\ref{fig:piprogen} denotes $X_{\rm max}$. 
We set the rotation velocity to 0 to match the condition with the 1D calculation and added a perturbation that uniformly increases the internal energy by $1\%$ at the start of calculation so that the maximum central temperature and density between the 1D and 2D calculations match.

\begin{figure*}[htpb]
	\begin{minipage}{0.5\hsize}
		\begin{center}
			\hspace*{-5em}  
			\includegraphics[angle=0,scale=0.7]{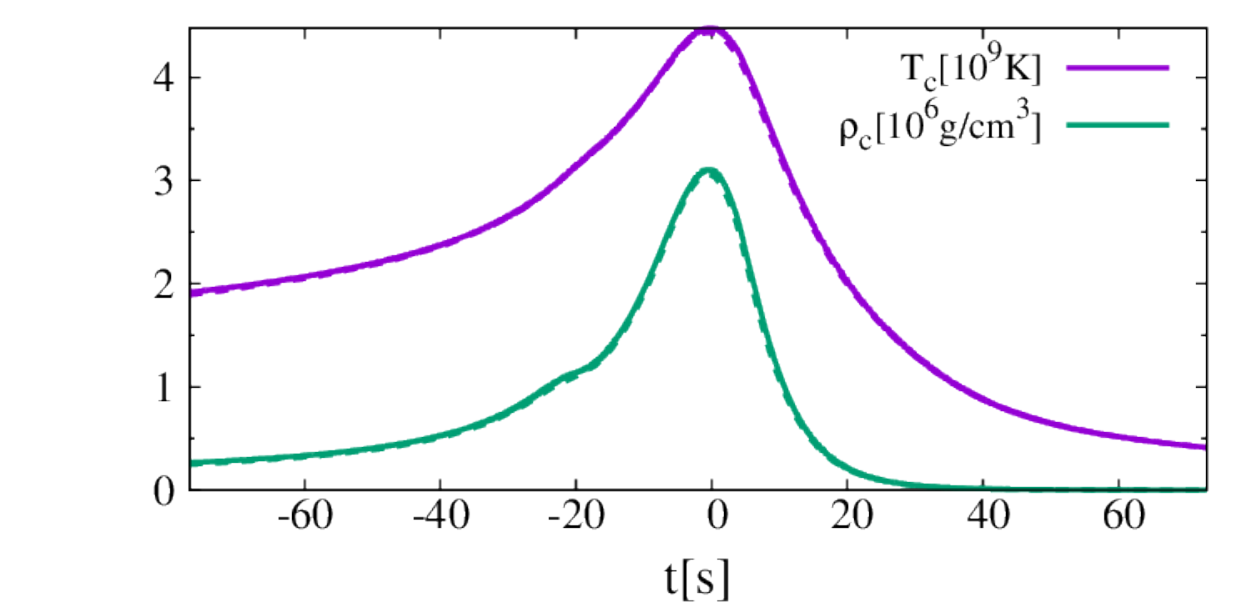}
		\end{center}
	\end{minipage}
	\begin{minipage}{0.5\hsize}
		\begin{center}
			\hspace*{-5em} 
			\includegraphics[angle=0,scale=0.7]{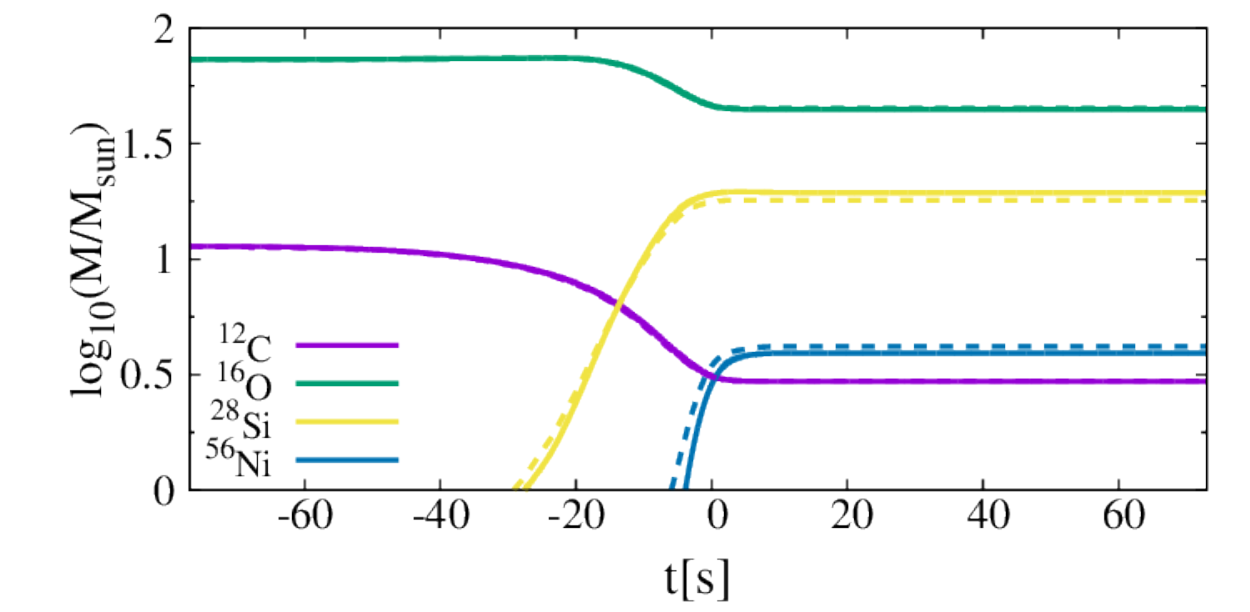}
		\end{center}
	\end{minipage}
	\caption{Time evolution of the central temperature and density (left panel) and total mass of each major element in the CO core (right panel). The origin of time is taken at the time at which the central density becomes the maximum value for each calculation. The solid and dashed curves show the results of 2D and 1D calculations, respectively.}
	\label{fig:m200}
\end{figure*}
Figure~\ref{fig:m200} shows the time evolution of central density and temperature (left panel) and total mass of 
each major element (right panel) in the CO core. The solid and dashed curves show the results of the 2D and 1D calculations, respectively. 
The origin of time is taken at the time at which  the central density becomes the maximum value for each calculation. 
Figure~\ref{fig:m200} shows that the results of the 1D and 2D calculations are in good agreement. 
This indicates that our 2D code would be able to handle the gravitational collapse and explosion by the nuclear reaction as accurately as the 1D code.

\end{document}